\documentclass[prd,reprint,floatfix,nobibnotes,nofootinbib,a4paper,showpacs,amsmath]{revtex4-1}
\usepackage{mathbbol}
\usepackage[dvips]{graphicx}
\newcommand*\mycirc[1]{\mbox{\textcircled{\protect\raisebox {-0.25ex}{#1}}}}
\begin{document}
\title{Charmonium spectroscopy and mixing with light quark and open charm states\\[.1cm]
from $\mathbf{n_F=2}$ lattice QCD}
\author{Gunnar S.\ Bali}
\email{gunnar.bali@ur.de}
\author{Sara Collins} 
\author{Christian Ehmann}
\affiliation{Institut f\"ur Theoretische Physik, Universit\"at Regensburg,
93040 Regensburg, Germany}
\date{\today}
\begin{abstract}
We study the charmonium spectrum including higher spin
and gluonic excitations. We determine an upper limit on the
mixing of the $\eta_c$ ground state with light pseudoscalar flavour-singlet
mesons and investigate the mixing of charmonia near open charm
thresholds with pairs of (excited) $D$ and $\overline{D}$ mesons.
For charm and light valence quarks and $n_{\mathrm{F}}=2$ sea quarks, we employ
the non-perturbatively improved Sheikholeslami-Wohlert
(clover) action. Excited states are accessed using the variational
technique, starting from a basis of suitably optimised operators.
For some aspects of this study,
the use of improved stochastic all-to-all
propagators was essential.
\end{abstract}
\pacs{12.38.Gc, 14.40.Pq, 12.39.Mk, 13.25.Gv}
\maketitle

\section{Introduction}

During the past decade several new charmonium resonances
were discovered, primarily by experiments at the
two $B$-factories but also by CLEO-c and at the Tevatron.
With BES-III and the LHC collecting data, possible Super-$B$ factories
and the planned PANDA experiment~\cite{Lehmann:2009dx} at the FAIR facility,
experimental prospects to study these states
in more detail and to discover further resonances are very promising. 
For an overview, see e.g.
Refs.~\cite{Brambilla:2010cs,Braaten:2008nv,Godfrey:2008nc,Barnes:2007zza,Voloshin:2007dx,Eichten:2007qx,Swanson:2006st}.

Current phenomenological debates
focus on the $X$, $Y$ and $Z$ resonances that are close
to or above open charm thresholds.
At least four different
frameworks have been suggested to accommodate
these states:
\begin{itemize}
\item $D^{(*)}\overline{D}$ molecules (or deusons)~\cite{De Rujula:1976qd,Novikov:1977dq,Tornqvist:1993ng,Swanson:2006st,Close:2010wq,Oset:2011bq}, composed of
a charmed meson $D^{(*)}$ and antimeson $\overline{D}$,
\item tetraquark states (or baryonia)~\cite{Jaffe:1976ig,Weinstein:1982gc,Manohar:1992nd,Maiani:2004vq,Maiani:2007vr} consisting of diquark-antidiquark
pairs, bound by QCD forces,
\item $\bar{c}cg$ hybrid (or hermaphrodite) states~\cite{Barnes:1977hg,Barnes:1982zs,Chanowitz:1982qj,Isgur:1985vy} consisting of a charm-anticharm
quark pair and additional gluons, and
\item a compact $\bar{c}c$ core, bound 
inside a light meson, hadro-charmonium~\cite{Voloshin:1976ap,Dubynskiy:2008mq}.
\end{itemize} 
One example of a molecule or tetraquark
candidate is the $X(3872)$. The $Y(4260)$ can
at present be accommodated as a hybrid or as
a hadro-charmonium state while the $Z^+(4430)$ (if confirmed)
could either be a molecule or hadro-charmonium.

The new states also pose novel challenges to lattice simulations.
In the case of standard charmonia that can be classified
according to a non-relativistic quark model, the 
sizeable quark mass $m_c>\Lambda$, where $\Lambda$ denotes
a typical hadronic binding energy,
represents the main difficulty:
lattice
artefacts, that in our case are of $\mathcal{O}[(m_ca)^2]$,
are usually not
small at currently available lattice spacings $a$.
In the $\Upsilon$ case 
the $b$ quark mass can be integrated out and an
effective field theory, non-relativistic QCD (NRQCD), simulated on the
lattice~\cite{Lepage:1992tx,Davies:1994mp}.
However, the charm quark mass $m_c$ is
not sufficiently large to allow for this. In this case
higher order perturbative or non-perturbative
corrections will be sizeable. Therefore the charm quark needs
to be simulated using a relativistic action.

One would expect observables that are very sensitive to
the mass $m_c$ to be more strongly affected by lattice artefacts
than those that are insensitive
to the precise value of this mass.
Using effective field theory methods like the Fermilab
approach
to heavy quarks~\cite{Kronfeld:2000ck,Harada:2001ei,Burch:2009az}
or NRQCD~\cite{Caswell:1985ui} and
potential NRQCD (pNRQCD)~\cite{Pineda:1997bj,Brambilla:1999xf}
some insight
can be gained into this. For instance, charm
quark mass effects on spin-averaged splittings
are suppressed by a factor of
the squared average relative velocity
of the charm quarks $v^2$.
Momentum exchanges $\propto m_cv$ 
in turn become relevant for the finestructure.
Finally, lattice spacing effects on
determinations of the mass parameter $m_c$ from the
charmonium spectrum are not suppressed by any powers of $v$.
This means that a computation of
the spin-averaged spectrum will be less demanding
with respect to the continuum limit extrapolation
than predictions of the finestructure or of the charm quark mass.

The standard spectroscopy of charmonium states including the
continuum limit extrapolation is well understood in the
quenched approximation to QCD, see e.g.~\cite{Bali:2006xt} and references
therein, and several new results
including sea quark flavours exist, on isotropic lattices~\cite{Follana:2006rc,Namekawa:2008ft,Namekawa:2011wt,Burch:2009az,Bali:2011dc}
as well as on anisotropic lattices~\cite{Ryan:2010zz} that employ a smaller
temporal than spatial resolution $a_t\ll a_s$, to lessen the
severity of the scale separation $m_cv>m_cv^2$.

An accurate reproduction of the charmonium finestructure
in the continuum limit represents an important test of
QCD and of lattice methods. However,
taking the continuum limit may be less vital to reproduce
qualitative features of loosely
bound open charm threshold states that are spatially
more extended. In this case one needs
to consider the mixing of states created by two-quark and
by four-quark operators.
Some pioneering studies have already been done, creating
states with $\bar{c}q\bar{q}c$ operators~\cite{Chiu:2005ey,Chiu:2006hd}
where $q$ denotes a light quark flavour.
However, so far disconnected quark loop diagrams and hence
annihilation channels have been neglected.
Moreover, lattice studies of the light quark sector, see e.g.\
Ref.~\cite{Prelovsek:2010kg}, and
of string breaking in the static limit~\cite{Bali:2005fu}
teach us that these diagrams
and, in particular, mixing between states created by $\bar{c}c$
and $\bar{c}q\bar{q}c$
operators can be important.

Here we will explore methods needed to systematically
study charmonium threshold states and apply these to phenomenology.
This article is organized as follows. In Sec.~\ref{sec:methods}
we will introduce
our methods, namely the gauge ensembles that are being used,
the smeared operators
that enter our variational analysis and
the all-to-all propagator techniques.
In Sec.~\ref{sec:spectrum} we
present results on standard charmonium spectroscopy at a finite
lattice spacing, employing two-quark
($\bar{c}c$) creation operators only. This includes higher spin
and exotic states and provides us with the improved operators
that are needed in Secs.~\ref{sec:etamix} and \ref{sec:moleculemix}.
In Sec.~\ref{sec:etamix} we investigate the mixing between
$\bar{c}c$ and $\bar{q}q$ operators. This will yield an upper limit
to the mixing between the flavour-singlet $\eta_c$ and
$\eta$ mesons that in principle could have an effect on the
$S$-wave finestructure. Finally, in Sec.~\ref{sec:moleculemix}
we investigate
the contribution of
four-quark $\bar{c}c\bar{q}q$ components to radially excited
charmonium states, before we
conclude in Sec.~\ref{sec:sum}.
Some preliminary results were presented at lattice
conferences~\cite{Ehmann:2007hj,Ehmann:2009ki,Bali:2009er}.

\section{Simulation details and methods}
\label{sec:methods}
\subsection{Gauge configurations}
\label{sec:gauge}
We base our study on $n_{\mathrm{F}}=2$ configurations
of the QCDSF Collaboration
generated using the non-perturbatively improved
Sheikholeslami-Wohlert
(clover) Fermion action~\cite{Sheikholeslami:1985ij}
and the Wilson gauge action, provided by the QCDSF
collaboration. Details can be found in Ref.~\cite{Ali Khan:2003cu}.
The charm quark mass $m_c$ is not sufficiently heavy to allow
for a non-relativistic action with controllable systematics.
Therefore, we use the same action for the charm quark as for
the light sea/valence quarks, with a well-defined $\mathcal{O}(a)$ improved
continuum limit. Note that except for the value of the
coefficient $c_{\mathrm{SW}}=c_B=c_E$ the clover action
is identical to the version of the Fermilab heavy quark action
used, e.g., in Ref.~\cite{Burch:2009az} and our results
at a finite lattice spacing $a$ may be interpreted accordingly.

\begingroup
\squeezetable
\begin{table}
\caption{Identifier (ID), simulation parameters, charm quark
$\kappa$-value ($\kappa_{\mathrm{charm}}$) and the number of analysed effectively
statistically independent gauge configurations of our
runs.\label{confdetail_tab}}
\begin{ruledtabular}
\begin{tabular}{ccccccccc}
 ID & $\beta$ & $\kappa$ & volume & $m_{\mathrm{PS}}/$GeV & $a/$fm & $L/$fm & $\kappa_{\mathrm{charm}}$ & $N_\mathrm{conf}$ \\
 \hline
 \mycirc{1} & 5.20 & 0.13420 & $16^3\times32$ & 1.007(2) & 0.1145 & 1.83 & 0.1163 & 100 \\
 \mycirc{2} & 5.29 & 0.13620 & $24^3\times48$ & 0.400(1) & 0.0770 & 1.84 & 0.1245 & 130 \\
 \mycirc{3} & 5.29 & 0.13632 & $24^3\times48$ & 0.280(1) & 0.0767 & 1.84 & 0.1244 & 100
\end{tabular} 
\end{ruledtabular}
\end{table}
\endgroup

We list the ensembles that we employ in
Table~\ref{confdetail_tab}, together with an identifier.
The lattice spacing is set from the value
$r_0\approx 0.467$~fm. With this choice
the nucleon mass agrees with experiment when extrapolated
to the physical light pseudoscalar mass\footnote{A recent
re-analysis yields somewhat different $r_0/a$-values~\cite{najjar},
in particular at small quark masses. Here we ignore these
developments. Otherwise we would have to rerun all simulations,
re-adjusting the charm quark mass by
$-5\,\%$, $-6\,\%$ and $+8\,\%$, on
ensembles \mycirc{1}, \mycirc{2} and \mycirc{3}, respectively. However,
most of the charmonium mass is given by
$2m_ca$ so that only mass splittings will be affected
by such a re-adjustment. Fortunately, the spin-averaged
splittings were found to be rather insensitive to the charm quark
mass~\cite{Bali:1998pi} while the main
systematics regarding the finestructure are the
unrealistic sea quark content and the missing
continuum limit extrapolation.}~\cite{Ali Khan:2003cu},
$m_{\mathrm{PS}}=m_{\pi}^{\mathrm{phys}}$.
The measured values of $r_0/a$ not only depend
on the inverse lattice coupling
$\beta$ but also on the mass parameter $\kappa$.
One can now decide to define a lattice spacing
$a(\beta,\kappa)$ or replace this by a chirally
extrapolated $a(\beta)$. 
After performing a chiral extrapolation in the sea
quark mass, the results of the two choices
obviously should agree for physical observables. Since in this
exploratory study we do not
attempt such an extrapolation, we decide to set the lattice spacing
from the $r_0/a(\beta,\kappa)$ values, as determined at the investigated
sea quark $\kappa$ parameters.

This leaves us with the charm quark
mass as the only free parameter which we set
by tuning,
\begin{equation}
\label{eq:sa}
m_{1\overline{S}}=\frac{1}{4}\left(m_{\eta_c}+3m_{J/\Psi}\right)\,,
\end{equation}
to its experimental value
of~\cite{Nakamura:2010zzi}, $(3067.8\pm 0.4)$~MeV.

The ensembles \mycirc{1} and \mycirc{2} are used 
to optimise the smearing functions.
Our study of mixing between the $\eta_c$ and the
light quark $\eta$-meson is performed on \mycirc{1} where the mass
gap between these states is smallest so that one may expect the
biggest effect. For the mixing with threshold states ensemble
\mycirc{3} is used because in this case
light $D$-meson masses are mandatory.

\subsection{The variational method}
\label{Se:varMeth}
We extract energy levels $E_n$ from the decay of two-point Green functions
in Euclidean time,
\begin{align}
C_{ij}(t)&=\langle O_i(t)O_j^{\dagger}(0)\rangle\label{eq:expect}\\
&=\sum_{n\geq 1} v_i^nv_j^{n*}e^{-E_nt}\,,\label{eq:expect2}
\end{align}
where $v_i^n=\langle 0|\hat{O}_i|n\rangle$.
In the case of the clover action link reflection
positivity is violated and so in principle the
above spectral decomposition with positive real energy eigenvalues
only becomes valid for sufficiently
large Euclidean times. In practice for
$t\geq a$ we do not detect any violations.
$\hat{O}_i^{\dagger}$ are operators creating states
of an isospin $I$, charm number\footnote{Here we do not consider
strangeness, beauty etc..} $C$, a given momentum and
$\mathrm{SO}(3)\,\otimes\,{\mathbb Z}_2\,(\otimes\,\,{\mathbb Z}_2)$ $J^{P(C)}$
quantum numbers.
Note that on the lattice the infinite dimensional
$\mathrm{O}(3)$ group is broken down to its
octahedral $\mathrm{O_h}$ subgroup of order 48, with only ten
($A_1, A_2, E, T_1$ and $T_2$ times parity)
irreducible Bosonic representations.
The mapping between the continuum $J$ spins and these $\mathrm{O_h}$ spins
is given in Tables~\ref{tab:lat1} -- \ref{tab:lat2}.

\begin{table}
\caption{Continuum spins that contribute to a given lattice
representation.\label{tab:lat1}}
\begin{ruledtabular}
\begin{tabular}{ccc}
irrep.&dimension&continuum $J$\\\hline
$A_1$&1&0,4,\ldots\\
$A_2$&1&3,\ldots\\
$E$&2&2,4,\ldots\\
$T_1$&3&1,3,4,\ldots\\
$T_2$&3&2,3,4,\ldots
\end{tabular}
\end{ruledtabular}
\end{table}

\begin{table}
\caption{The ``inverse'' of Table~\protect{\ref{tab:lat1}}.
Lattice spins that are
``embedded'' within each continuum spin.\label{tab:lat2}}
\begin{ruledtabular}
\begin{tabular}{ccc}
$J$&$\mathrm{O_h}$ rep.&dimensions\\\hline
0&$A_1$&1\\
1&$T_1$&3\\
2&$E$, $T_2$&2+3\\
3&$A_2$, $T_1$, $T_2$&1+3+3\\
4&$A_1$, $E$, $T_1$, $T_2$&1+2+3+3\\
$\cdots$&$\cdots$&$\cdots$
\end{tabular}
\end{ruledtabular}
\end{table}

The expectation value Eq.~(\ref{eq:expect})
will depend on the time difference between
creation and destruction of the state so that for convenience we
have set the source time to zero.
Obviously, $C_{ij}(t)=C_{ji}^*(t)$ is Hermitian and in our case we will use
operators with phases so that it is real and positive definite
for $t\geq a$. For large times $t$ the exponential
decay of the $C_{ij}(t)$ entries will be governed by the
ground state energy $E_1$, or, for a momentum
${\mathbf p}={\mathbf 0}$, by the ground state mass.
Due to the translational invariance of
the expectation value, it is sufficient to perform this
momentum projection at the sink. 
We do this for the standard spectroscopy
so that we only need to generate point-to-all propagators
in this case. Note that we still have the symmetry
$C_{ij}(t)=C_{ji}(t)$ in the limit of infinite statistics, however,
the statistical errors of $C_{ij}(t)$ and $C_{ji}(t)$ for $i\neq j$ 
will not be of similar sizes.
Replacing off-diagonal elements so 
that more smearing iterations (see Sec.~\ref{sec:fermsmear})
are applied at the source
than at the (momentum-projected) sink reduces the statistical errors.

The convergence in Euclidean time of effective masses,
\begin{equation}
m_{ij,{\mathrm{eff}}}(t+a/2)=a^{-1}\ln\frac{C_{ij}(t)}{C_{ij}(t+a)}\,,
\end{equation}
towards the ground state mass is affected by the quality
of the ground state
overlap $c_i=|v_i^1|^2=|\langle 1|\hat{O}_i^{\dagger}|0\rangle|^2$ of the operator
$\hat{O}_i$. Having many different such operators at our disposal
enables us not only to determine the ground state energy at small
$t$-values where statistical errors are small but also
allow us to access excited states, using the variational
approach~\cite{Michael:1985ne,Luscher:1990ck,Blossier:2009kd},
also known as the generalized eigenvalue approach.

We choose a basis of operators $\hat{O}_i$, $i=1,\ldots,N$, destroying
a colour singlet state within a given lattice
representation. These operators may differ for example
by their spatial extents or their Fock structures and they are usually
not mutually orthogonal.
These are then used to construct the correlation matrix Eq.~(\ref{eq:expect}).
We now solve the symmetrized eigenvalue problem,
\begin{equation}
\label{evp} 
 C^{-1/2}(t_0)C(t)C^{-1/2}(t_0)\psi^n(t,t_0) = \lambda^n(t,t_0)
 \psi^n(t,t_0)\,.
\end{equation}
Note that
$C^{-1/2}(t_0)C(t)C^{-1/2}(t_0)={\mathbb 1}$ at the normalization
time $t=t_0$: everything
is expressed relative to the eigenbasis of $C(t_0)$.
We order $\lambda^1(t)>\lambda^2(t)>\cdots>\lambda^N(t)>0$ at large $t$.
To ensure consistency over jackknife samples, in the statistical
analysis we also monitor
the directions of the eigenvectors.
Note that the original non-symmetrized definition of Ref.~\cite{Michael:1985ne} yields the same eigenvalues but different, non-orthogonal eigenvectors,
$\phi^n(t,t_0)=C^{-\frac12}(t_0)\psi^n(t,t_0)$,
\begin{equation}
C^{-1}(t_0)C(t)\phi^n(t,t_0)=\lambda^n(t,t_0)\phi^n(t,t_0)\,.
\end{equation}

If we choose $t_0$ overly large then excited states will have died out in
Euclidean time and the rank of $C(t_0)$
will not be maximal, within the given statistical errors.
For $t_0$ chosen too small,
$C(t)$ will receive contributions from more than the $N$ lowest
lying states, resulting in unstable eigenvectors and eigenvalues.
It can be shown that the eigenvalues behave like~\cite{Blossier:2009kd}, 
\begin{equation}
\label{varmeth_ev}
  \lambda^n(t,t_0) \propto e^{-(t-t_0) \, E_n} [1+{\mathcal O}(e^{-(t-t_0) \, \Delta E_n})] \,,
\end{equation}
where $\Delta E_n$ is the energy difference between the energy of the
first state not contained in the operator basis\footnote{At least to first
order in perturbation theory. To second order states with energies
$\leq E_n$ can contribute as well, at $t\gg t_0$.
In Ref.~\cite{Blossier:2009kd} it has been shown that
these effects are negligible for $t\leq 2t_0$. In general
the maximum admissible value of $t$ at a given $t_0$ depends on
the underlying spectrum and on the basis of trial wavefunctions
used.}, $E_{N+1}$ and $E_n$.
The correction factor arises from the finite dimensionality
and non-orthogonality of the operator basis. Ideally one will aim at a
set of operators that dominantly couple to the first $N$ states and
that are as orthogonal as possible to each other.
The eigenvectors, up to the rotation
and the change in the normalization of Eq.~(\ref{evp}),
approach their physical counterparts
$v^n$ of Eq.~(\ref{eq:expect2}) too, with
similar exponential corrections in Euclidean time~\cite{Blossier:2009kd}.

From the eigenvalues we can also define effective energy levels,
or, for ${\mathbf p}={\mathbf 0}$, masses,
\begin{equation}
m_{n,\mathrm{eff}}^{t_0}(t+a/2)
=a^{-1}\ln\frac{\lambda^n(t,t_0)}{\lambda^n(t+a,t_0)}\,,
\end{equation}
that, for sufficiently large $t_0$ and $t>t_0$, should exhibit
plateaus which we then fit to a constant to obtain the masses $m_n$.

\begin{figure}[ht]
\includegraphics[height=.48\textwidth,clip,angle=270]{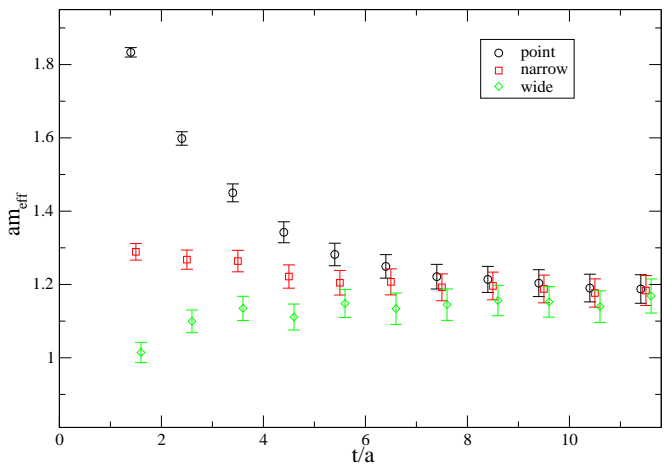}
 \caption{Effective masses of correlation functions between a
point source and point, narrow and wide smeared sinks.
\label{effmass_fig}}
\end{figure}
\subsection{Fermion field smearing}
\label{sec:fermsmear}
The variational method is the central tool of our analysis.
It needs to be supplied with suitable building blocks in terms
of operators, from which good approximations of the physical
eigenstates can be obtained. The wavefunctions of physical
eigenstates will not be ultra-local objects and
spatially extended interpolators need to be considered.
We generate such
extended operators by applying 
Wuppertal smearing~\cite{Gusken:1989qx} to a Fermion field $\psi$, 
\begin{equation}
\label{wupp}
\psi^{(n)}_x=\frac{1}{1+6\delta}\left(\psi^{(n-1)}_x+
\delta\sum_{j=\pm 1}^{\pm 3}\overline{U}_{x,j}\psi^{(n-1)}_{x+a\hat{\boldsymbol{\jmath}}}\right)\,.
\end{equation}
$n=1,\ldots, n_{\mathrm{wup}}$ counts the iteration number and
$\delta>0$ is a free parameter.
The (arbitrary) normalization
convention is chosen to avoid numerical overflows
for large iteration counts
$n_{\mathrm{wup}}$. $\overline{U}_{x,j}$ is a gauge
covariant transporter, connecting the lattice point $x$ with its spatial
neighbour in the $j$-direction,
$x+a\hat{\boldsymbol{\jmath}}$, for instance a gauge link $U_{x,j}$.
In our implementation we used APE smeared~\cite{Falcioni:1984ei}
links for $\overline{U}_{x,j}$,
see Sec.~\ref{sec:ape} below.
Note that the Wuppertal smearing operator is gauge covariant. It
transforms as a singlet under $\mathrm{O_h}$, parity and charge transformations,
it is Hermitian, translationally invariant and spin-diagonal.

\begin{figure}[ht]
\includegraphics[width=.48\textwidth,clip]{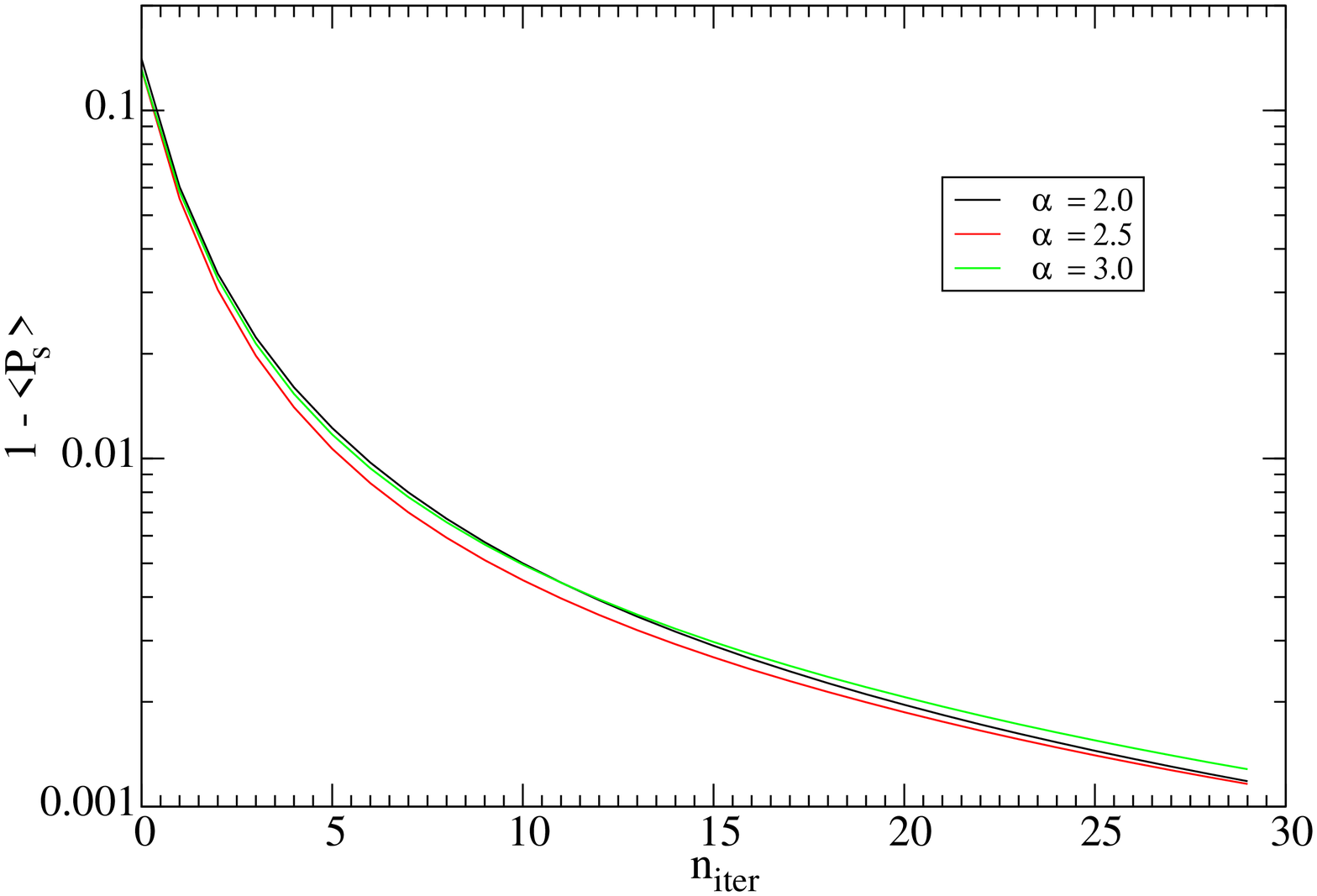}
 \caption{Deviations of the average spatial plaquette from unity,
against the number of APE
smearing iterations Eq.~(\protect\ref{eq:smear}) on a
lattice of ensemble \mycirc{1} for different $\alpha$ values.
\label{ape}}
\end{figure}

We can rewrite the above equation by defining a covariant
spatial lattice Laplacian,
\begin{equation}
a^2\left(\nabla^2\psi\right)_x=-6\psi_x
+\sum_{j=\pm 1}^{\pm 3}\overline{U}_{x,j}
\psi_{x+a\hat{\boldsymbol{\jmath}}}\,,\end{equation}
to obtain,
\begin{equation}
\psi^{(n)}=\psi^{(n-1)}+\frac{\delta}{1+6\delta}\,a^2\nabla^2\psi^{(n-1)}\,.
\end{equation}

We introduce a fictitious time $t=n_{\mathrm{wup}}\Delta t$,
\begin{equation}
\label{eq:diffusion}
\frac{\partial\psi(t)}{\partial t}\approx
\frac{\psi(t+\Delta t)-\psi(t)}{\Delta t}
=k\frac{a^2}{\Delta t}\nabla^2\psi(t)\,,\end{equation}
where
\begin{equation}
k=
\frac{\delta}{1+6\delta}\,.\end{equation}
The diffusion equation Eq.~(\ref{eq:diffusion}) is formally solved by,
\begin{equation}
\psi(t)\approx e^{k (t/\Delta t)a^2\nabla^2}\psi(0)\,.\end{equation}
Starting from a $\delta$-source
$\psi_x(0)=\delta_{x0}$ on a free configuration
$U_{x,j}={\mathbb 1}$ this results in a Gauss packet
with the root mean square (rms) radius
of $\psi^{\dagger}\psi$,
\begin{equation}
\label{eq:radius}
\frac{\Delta r}{a}
=3\sqrt{k t/\Delta t}=3\sqrt{\frac{\delta}{1+6\delta}}\,\sqrt{n_{\mathrm{wup}}}\,.\end{equation}
The diffusion speed is maximal for $\delta\rightarrow\infty$
($k\rightarrow 1/6$) while the resulting wavefunction
is more continuum-like for $\delta\rightarrow 0$ ($k\rightarrow 0$).
As a compromise we choose $\delta=0.3$ ($k\approx 1/9.3$). 

\begin{figure*}[ht]
\includegraphics[width=.48\textwidth,clip]{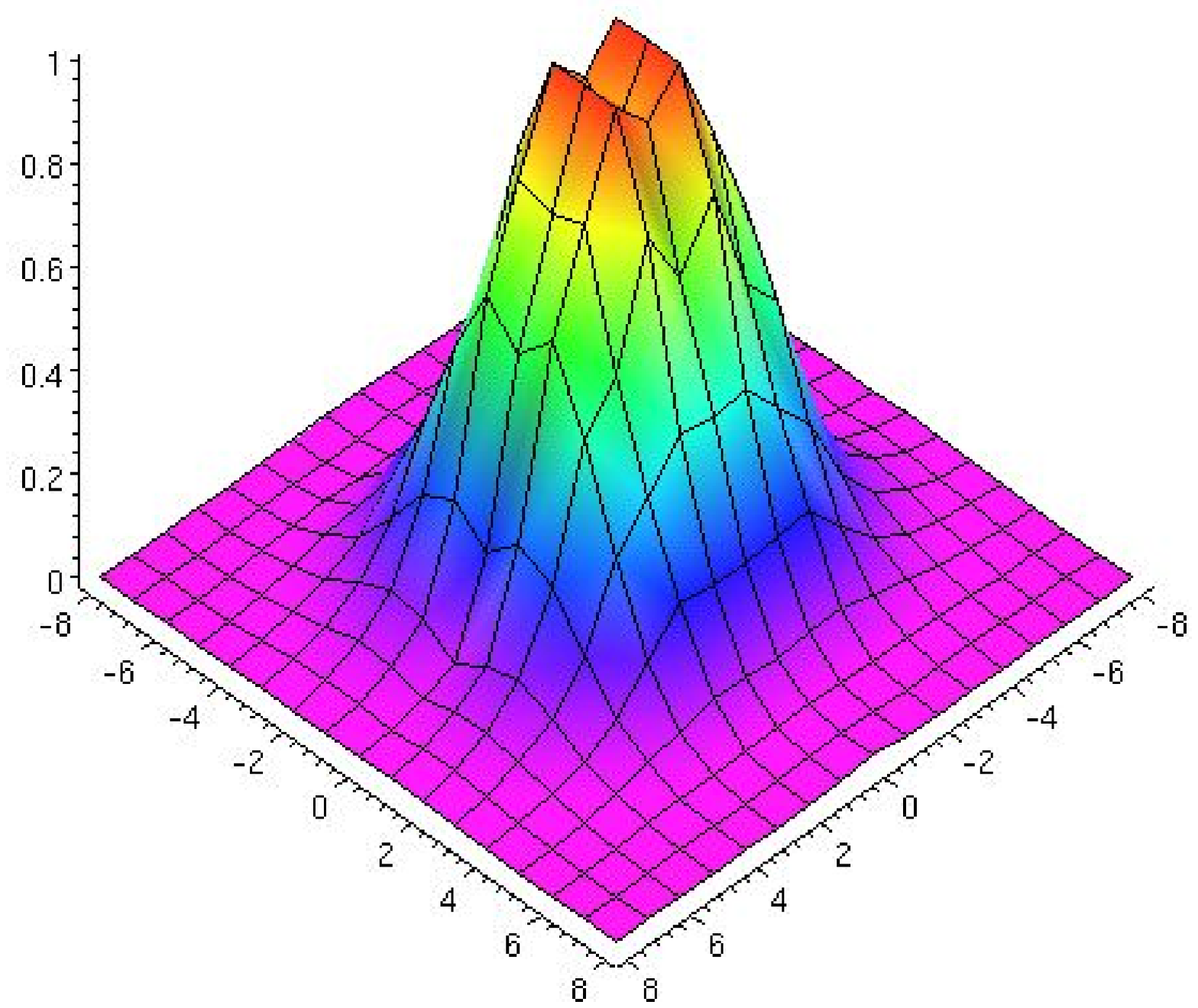}~
\includegraphics[width=.48\textwidth,clip]{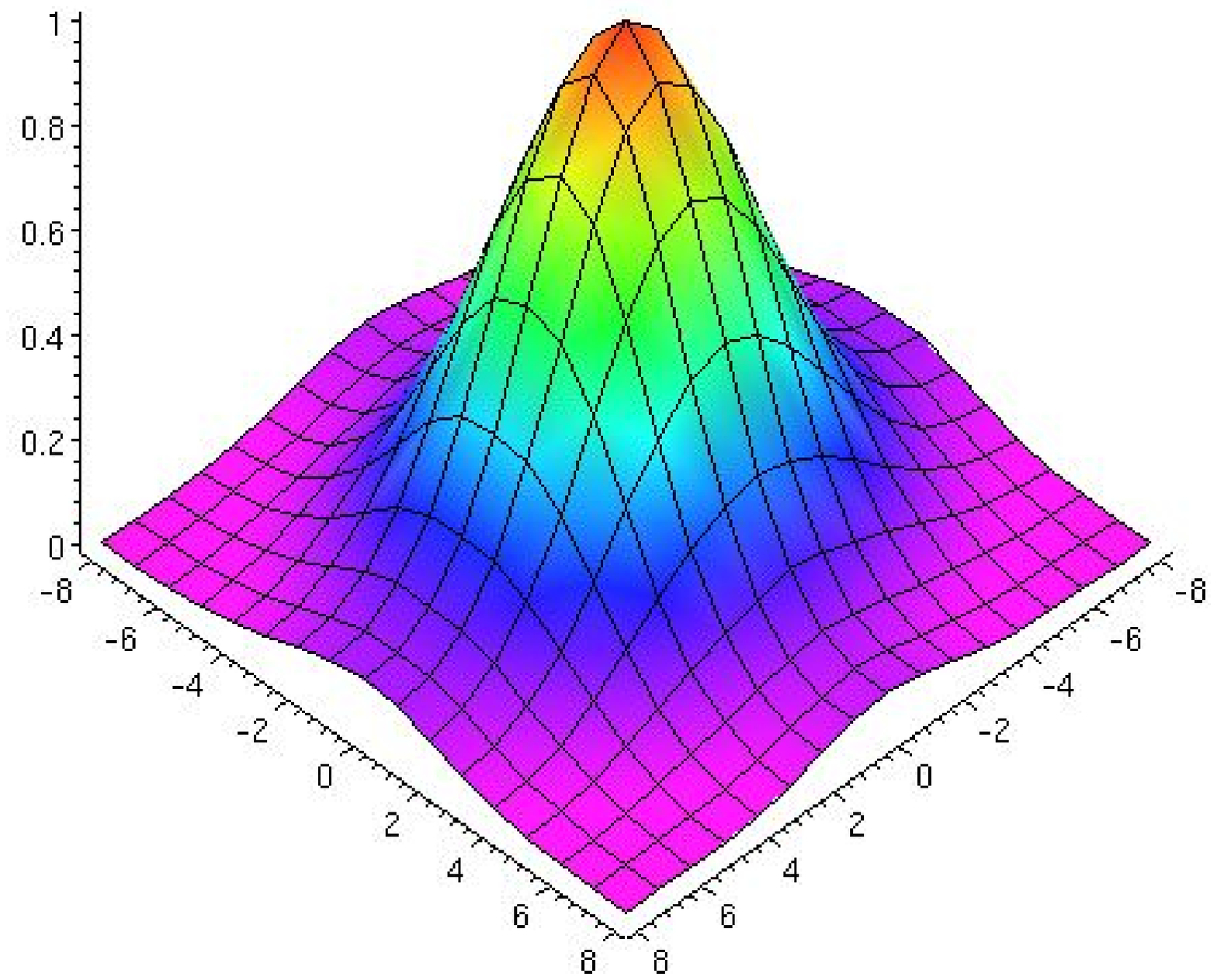}
\caption{The Wuppertal smearing function
($n_{\mathrm{wup}}=100$, $\delta=0.3$) with the original gauge links as
parallel transporters (left) and with APE smeared transporters (right).
\label{ape_wf}}
\end{figure*}

By adjusting $n_{\mathrm{wup}}$ we can control the overlap of our
trial wavefunctions with the physical states. Using a
point operator will lead to an effective mass with a significant
curvature at small Euclidean times. A few iterations of smearing
can help to flatten this out, suppressing the overlap with high excitations
that have many nodes in their wavefunctions.
Our strategy is to use an operator basis with a point operator
that couples well to excited states, a narrow operator that couples
well to the ground state and one operator that is somewhat wider.

In Fig.~\ref{effmass_fig} we display effective
masses for the pseudoscalar charmonium state,
with a $\bar{c}\gamma_5 c$ point source and with a point as
well as with smeared sinks,
on ensemble~\mycirc{2}, see Table~\ref{confdetail_tab}.
Note that since creation and destruction operators differ, in
the smeared cases the effective masses do not need to
decrease monotonically.
We employed $n_{\mathrm{wup}}=20$ and 80 smearing iterations for
the narrow and wide sinks, respectively. Note that we smeared both
quark and antiquark fields so that the effective radius of the
charmonium creation operator is by a factor $\sqrt{2}$ bigger than
the expectation in Eq.~(\ref{eq:radius}). 

\subsection{Gauge field smearing}
\label{sec:ape}
It was already suggested in Ref.~\cite{Gusken:1989qx} to replace
the gauge links within Eq.~(\ref{wupp}) by other covariant transporters
$\overline{U}_{x,j}$, that depend on spatial links within the
given timeslice. The ground state wavefunction is smooth and
so we may wish to reduce the gauge field fluctuations as well, to enhance
the overlap with low lying states. 
Following Ref.~\cite{Bali:2005fu} we employ
spatial APE smearing~\cite{Falcioni:1984ei} to the gauge links that enter
the Wuppertal smearing, iteratively replacing a link by a linear
combination of the link and the sum of the four surrounding spatial staples,
\begin{align}
\label{eq:smear}
V_{x,i}^{(n)} & =  U_{x,i}^{(n-1)}+\alpha\sum_{|j|\neq i}
U_{x,j}^{(n-1)}U^{(n-1)}_{x+a\hat{\boldsymbol{\jmath}},i}U^{(n-1)\dagger}_{x+a\hat{\boldsymbol{\imath}},j}\,, \nonumber \\
U_{x,i}^{(n)} & =   P_{\mathrm{SU}(3)} V_{x,i}^{(n)}\,.
\end{align}
$\alpha>0$ is a weight factor and $P_{\mathrm{SU}(3)}A$ projects $A$
onto $U\in \mathrm{SU}(3)$ so that
$\mathrm{Re} \mathrm{Tr}\,(UA^{\dagger})$ is maximal. This procedure
somewhat deviates from the definition of
Ref.~\cite{Bali:2005fu} but also preserves gauge covariance.

The spatial plaquette $\langle P_s\rangle$ measures the
curvature of the gauge fields. Maximizing this means a smoother
gauge background. In Fig.~\ref{ape},
$1-\langle P_s \rangle$ is plotted against the number of APE
smearing iterations on lattices of ensemble \mycirc{1}
(see Table~\ref{confdetail_tab})
for three
values of $\alpha$. The
approach to unity depends very little on
the gauge configuration or on the gauge ensemble that we use.
We decide to terminate the APE smearing after
$n_{\mathrm{ape}}=15$ iterations, using $\alpha=2.5$, as a compromise between
gauge field smoothness and the computer time spent.

APE smearing brings the links close to unity while preserving
the gauge covariance of the Wuppertal smearing. This means that
Eq.~(\ref{eq:radius}), which is valid for $\Delta r\gg a$ on
trivial gauge fields, is satisfied with good accuracy.
In Fig.~\ref{ape_wf} we display a colour component
after applying $n_{\mathrm{wup}}=100$ smearing iterations to a
$\delta$-source in Coulomb gauge, without and with APE smearing. 
Indeed, the trial wavefunction looks smoother and moreover,
we obtain the rms radius expected from Eq.~(\ref{eq:radius}).

Note that the APE smeared fields are only used to improve the operators
but not to propagate the quarks. The inversions of the lattice
Dirac operator are performed on the original gauge configurations.

\subsection{All-to-all propagators}
\label{Se:a2aprop}
We first introduce the stochastic method to estimate
all-to-all propagators. 
We then describe the improvement
methods that we employ, namely staggered spin partitioning
(SSP)~\cite{Ehmann:2009ki},
the hopping parameter expansion (HPE)~\cite{Thron:1997iy}
and recursive noise subtraction (RNS)~\cite{Ehmann:2009ki}.
We finally investigate the efficiency of combinations of
these methods for a realistic example.
Note that on top of this we also employ the
truncated solver method (TSM)~\cite{Collins:2007mh,Bali:2009hu},
see also Ref.~\cite{Alexandrou:2011ar},
that turns out to be beneficial even for masses
as heavy as that of the charm quark. We restrict its use to
light quark propagators though.

\subsubsection{Definitions and basics}
We denote the improved
lattice Wilson-Dirac operator by,
\begin{equation}
\label{eq:mwilson}
M=\frac{1}{2\kappa}\left({\mathbb 1}-\kappa D\right)\,.
\end{equation}
This will depend on the quark mass through $\kappa$.
For each of the 12 $\delta$-sources
$|0,\alpha,a\rangle$
at spacetime position $0$, spin $\alpha$ and colour
$a$ we can compute
solutions $|s^{0,\alpha,a}\rangle$
of the linear systems,
\begin{equation}
\label{eq:solve}
M|s^{0,\alpha,a}\rangle=|0,{\alpha},a\rangle\,.
\end{equation}
This defines the point-to-all propagator,
\begin{equation}
\label{eq:p2a}
S(x|0)^{ba}_{\beta\alpha}=s^{0,\alpha,a}(x,\beta,b)\,.
\end{equation}
Due to translational invariance of expectation values,
point-to-all propagators are often sufficient to calculate
hadronic two-point Green functions. However, if one had all-to-all
propagators at one's disposal, one would gain statistics from
self-averaging over different source points.
Moreover, some Wick contractions inevitably
lead to diagrams containing disconnected quark loops whose evaluations
require more than a few source points. 
Solving the 12 equations Eq.~(\ref{eq:p2a})
for all $V$
lattice points (in our case, $V=131072$ and 663552)
instead of for a single $x_0=0$
is prohibitive in terms of computer time and memory.

However, we encounter statistical
errors anyway from the path integral importance sampling
in the calculation of expectation values.
Hence it is sufficient to aim
at an unbiased estimate, which can be obtained using stochastic
methods~\cite{Bitar:1988bb,Dong:1991xb}.
We introduce the following notation,
\begin{equation}
\overline{A}=\overline{A}^N:=\frac{1}{N}\sum_{j=1}^NA^j\,,
\end{equation}
and define
random noise vectors $|\eta^j\rangle$, $j=1,\ldots,N$ with components,
\begin{equation}
\eta^j(x,\alpha,a)=\langle x,\alpha,a|\eta^j\rangle\in\frac{1}{\sqrt{2}}
\left({\mathbb Z}_2 \otimes i{\mathbb Z}_2\right)\,.
\end{equation}
This complex ${\mathbb Z}_2$ noise has the properties,
\begin{align}
\overline{|\eta\rangle}^N&= {\mathcal O}\left(1/\sqrt{N}\right)\,,\\
\overline{|\eta\rangle\langle\eta|}^N&= \mathbb{1}
+{\mathcal O}\left(1/\sqrt{N}\right)\,.\label{eq:cancel}
\end{align}
By solving,
\begin{equation}
M|s^j\rangle =|\eta^j\rangle\,,
\end{equation}
for $|s^j\rangle$, $j=1,\ldots,N$, one can construct an
unbiased estimate, see Eq.~(\ref{eq:cancel}),
\begin{align}
\label{eq:set}
M^{-1}_{\mathrm{E}}&:=\overline{|s\rangle\langle\eta|}\approx M^{-1}\,,\\
M^{-1}_{\mathrm{E}}&=M^{-1}-
M^{-1}\left({\mathbb 1}-\overline{|\eta\rangle\langle\eta|}\right)\,.
\label{eq:reduce}
\end{align}
${\mathbb 1}-\overline{|\eta\rangle\langle\eta|}$ is an off-diagonal matrix
with entries of ${\mathcal O}(1/\sqrt{N})$. Hence
the difference between the approximation of Eq.~(\ref{eq:set}) above
and the exact result reduces like $1/\sqrt{N}$.
When averaging over $n_{\mathrm{conf}}$ gauge configuration 
the additional stochastic errors of an estimated observable
reduce like $1/\sqrt{Nn_{\mathrm{conf}}}$
since the estimate is unbiased. One would like
to achieve some sort of balance where this stochastic error
becomes smaller than the unavoidable
gauge error $\propto 1/\sqrt{n_{\mathrm{conf}}}$
from averaging over the
configurations. Since the ratio of these sources of errors
is independent of $n_{\mathrm{conf}}$, once this is large enough for
the central limit theorem to hold, this optimization can
be performed on a small number of configurations.
Depending on the observable, a 
large number of estimates $N$ may be required,
unless the difference Eq.~(\ref{eq:reduce}) can be reduced.
Indeed, many methods of improving estimates exist,
see, e.g.~\cite{Bali:2009hu}
and references therein.

Below we introduce the improvement methods that we use
in this article.

\begin{figure}
\includegraphics[width=.48\textwidth,clip]{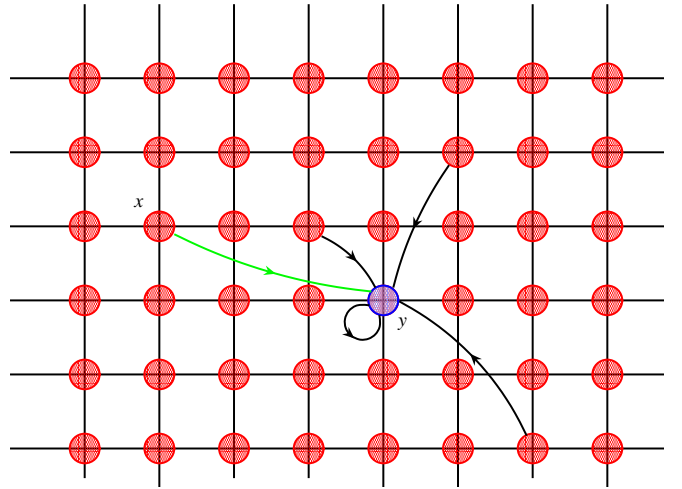}
 \caption{Two dimensional sketch of a global noise source. For the
propagator from $x$ to $y$ only the green line contributes
to the signal, the black ones contribute to the noise.}
\label{nodil}
\end{figure}
\subsubsection{Staggered spin partitioning}

One source of large uncertainties of the na\"{\i}ve estimate
is that the noise source vectors have entries on all lattice sites. 
The site, where the propagator ends, is surrounded by components
of the source vector that may not contribute to the signal but that will
contribute to the noise.
To see this more clearly, consider the estimation of a
propagator Eqs.~(\ref{eq:solve}) -- (\ref{eq:p2a})
connecting the point $x$ with the point $y$ [see also
Eq.~(\ref{eq:reduce})],
\begin{align}
S_E(y|x)_{\beta\alpha}^{ba}&=S(y|x)_{\beta\alpha}^{ba}\\\nonumber
&-\sum_{z,\gamma,c}S(y|z)_{\beta\gamma}^{bc}
\left[{\mathbb 1}-\overline{|\eta\rangle
\langle\eta|}\right]\!(z|x)_{\gamma\alpha}^{ca}\,,
\end{align}
where only entries with either $z\neq x$, $\gamma\neq\alpha$
or $c\neq a$ give non-vanishing
contributions to the noise sum, see Fig.~\ref{nodil}.

Since signals decrease exponentially with Euclidean distances,
\begin{equation}
\label{sigdecr}
||S(y|z)|| \sim e^{-|y-z|/\xi} \,,
\end{equation}
the source components located in the nearest neighbourhood of $y$
contribute most to the noise\footnote{This  heuristic argument is
over-simplistic since $S$ is not a gauge invariant quantity. However,
similar calculations can be performed for errors
of physical observables, with the same result.}
and thus it is desirable
to reduce or to remove these
terms entirely.
Likewise, contributions that are off-diagonal
in spin or colour at $y$ should be avoided. A brute force way
of achieving this
is by ``partitioning''~\cite{Bernardson:1993yg,Viehoff:1997wi,Wilcox:1999ab}.
For the special case of
spin partitioning (SP) this is
also known as the spin explicit method~\cite{Viehoff:1997wi}.

Partitioning amounts to decomposing
${\mathcal R}=V\,\otimes\,\mbox{colour}\,\otimes\,\mbox{spin}$
into $m$ subspaces ${\mathcal R}_j$:
${\mathcal R}=\oplus_{j=1}^m{\mathcal R_j}$.
One can set the source vectors $|\eta^i_{|j}\rangle$
to zero, outside of the subspace ${\mathcal R_j}$ and
label the corresponding solutions as $|s^i_{|j}\rangle$. 
The estimate of the all-to-all propagator is then given by
the sum, \begin{equation}
M^{-1}_E= \sum_{j=1}^m \overline{|s_{|j}\rangle\langle
\eta_{|j}|}\,.
\end{equation}
Clearly, this results in an $m$-fold increase of the total number of
solver applications.
If the stochastic noise reduction exceeds
a factor $1/\sqrt{m}$ then this
computational overhead is justified.

\begin{figure}
\large
\begin{center}
{\large
\begin{tabular}{c|c|c|c|c|c|c|c}
1 & 1 & 1 & 1 & 1 & 1 & 1 & 1 \\
\hline
1 & 1 & 1 & 1 & 1 & 1 & 1 & 1 \\
\hline
1 & 1 & 1 & 1 & 1 & 1 & 1 & 1 \\
\hline
1 & 1 & 1 & 1 & 1 & 1 & 1 & 1
\end{tabular}}
 \caption{Two dimensional schematic sketch of standard spin partitioning.
The numbers indicate the spinor component filled at the specific lattice site
for set 1 (out of 4).
\label{standspindil}}
\end{center}
\end{figure}

\begin{figure}
{\large
\begin{tabular}{c|c|c|c|c|c|c|c}
1 & 2 & 3 & 4 & 1 & 2 & 3 & 4 \\
\hline
2 & 3 & 4 & 1 & 2 & 3 & 4 & 1 \\
\hline
3 & 4 & 1 & 2 & 3 & 4 & 1 & 2 \\
\hline
4 & 1 & 2 & 3 & 4 & 1 & 2 & 3
\end{tabular}
\hspace*{.05\textwidth}
\begin{tabular}{c|c|c|c|c|c|c|c}
1 & 3 & 2 & 4 & 1 & 3 & 2 & 4 \\
\hline
3 & 2 & 4 & 1 & 3 & 2 & 4 & 1 \\
\hline
2 & 4 & 1 & 3 & 2 & 4 & 1 & 3 \\
\hline
4 & 1 & 3 & 2 & 4 & 1 & 3 & 2
\end{tabular}}
 \caption{Two dimensional schematic sketch of odSSP~(left) and obdSSP~(right).
The numbers indicate the spinor component filled at the specific lattice site
for set 1 (out of 4).
\label{stagspindil}}
\end{figure}

\begin{figure}
\parbox{.21\textwidth}{\includegraphics[width=.19\textwidth,clip]{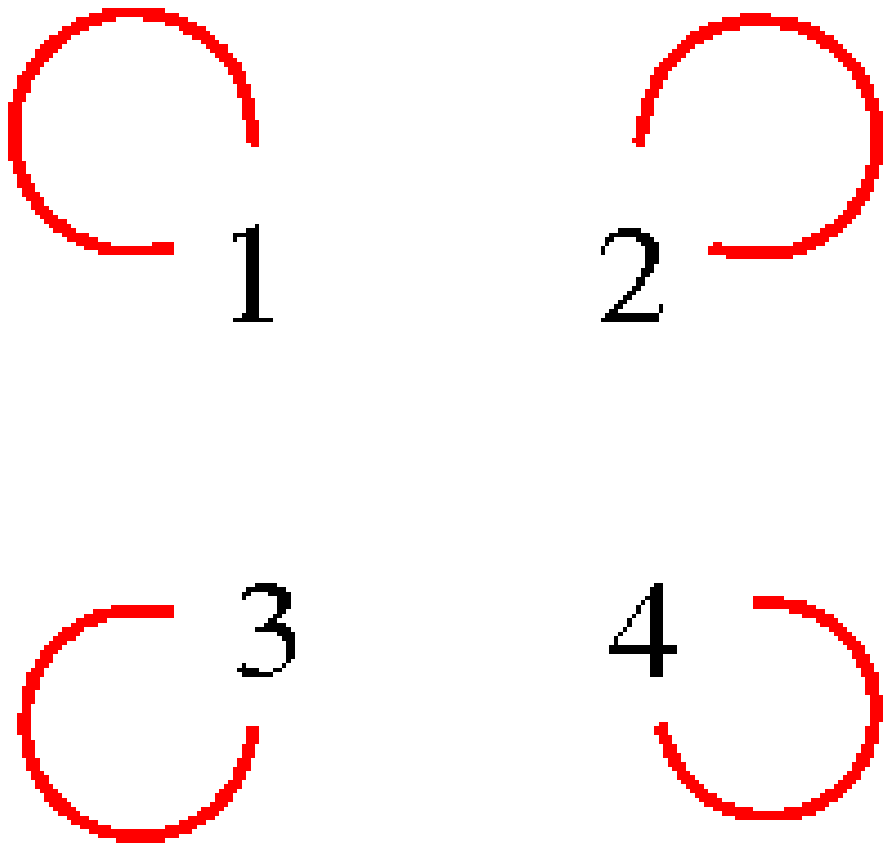}}\hspace*{.01\textwidth}
\parbox{.11\textwidth}{\includegraphics[width=.10\textwidth,clip]{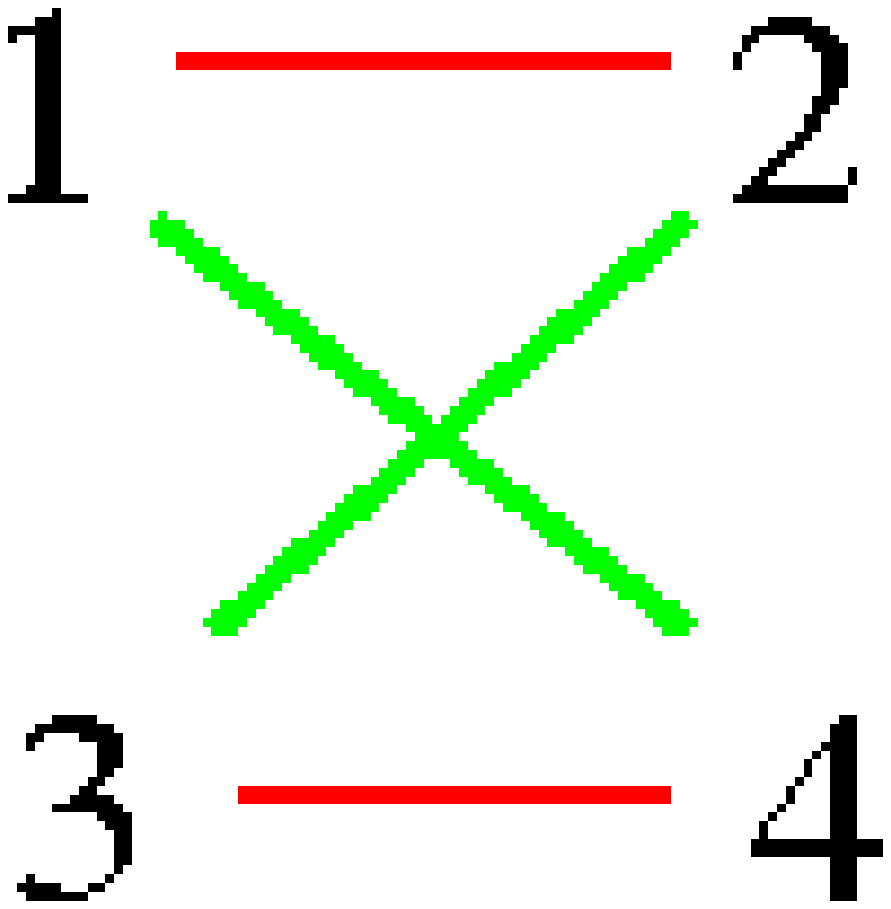}}\hspace*{.04\textwidth}
\parbox{.105\textwidth}{\includegraphics[width=.10\textwidth,clip]{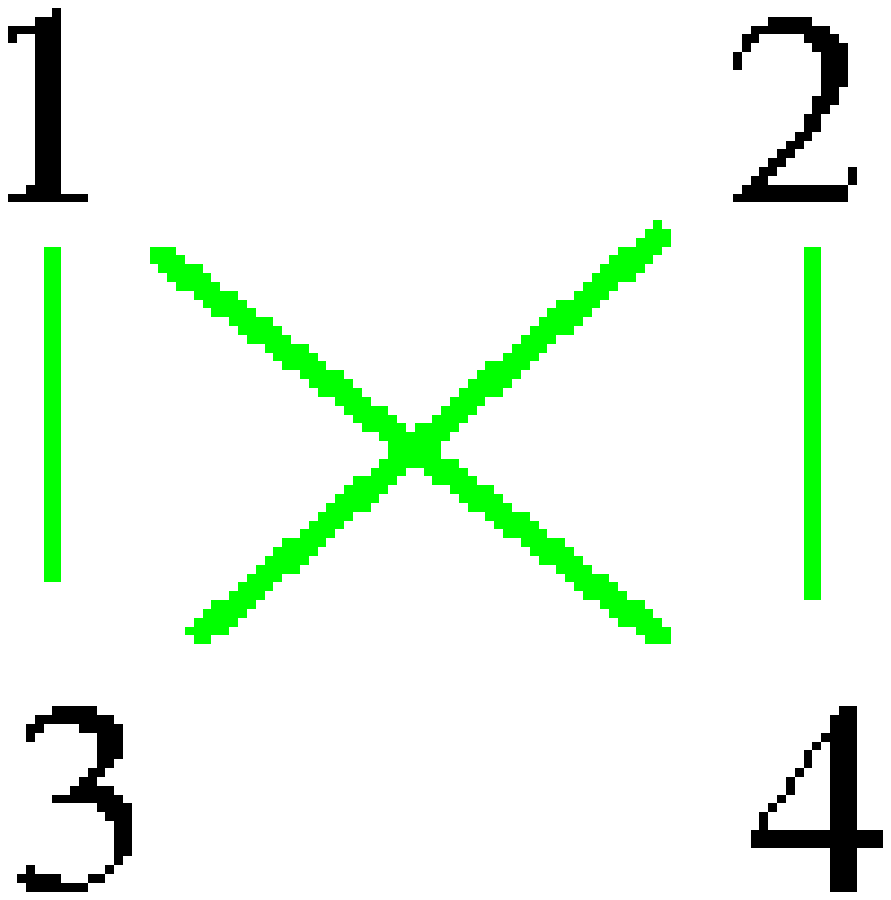}}
 \caption{The coupling strengths between the
spinor components of standard~(left), off-diagonal~(centre)
and off-block-diagonal~(right) spin partitionings.
Red indicates a strong (i.e.\ nearest neighbour) coupling,
green a weak (i.e.\ next to nearest neighbour) coupling.
\label{couplspindil}}
\end{figure}

Here we utilize spin and colour
partitioning. So far within each spin partitioning set
the same spinor component was dialled on every lattice site,
see Fig.~\ref{standspindil}.
Depending on the observable, however, it may be favourable to alter
the component to be filled within a specific set as a function of the
spacetime position. For heavy quarks, the coupling between the upper and the
lower two components of the Dirac spinor is small. One may exploit this
by separating in spacetime components that couple strongly,
only allowing for weakly coupled components to share a link.
We call such non-trivial spin partitioning schemes
staggered spin partitioning (SSP)~\cite{Ehmann:2009ki}.
In Fig.~\ref{stagspindil} we sketch two SSP versions,
off-diagonal SSP (odSSP) and off-block-diagonal SSP (obdSSP).
Fig.~\ref{couplspindil} illustrates the coupling strengths between
the four spinor components. Red lines indicate nearest neighbours
(strong coupling) and green lines next to nearest neighbours (weak coupling).

The obdSSP scheme should be particularly well suited to heavy quarks.
However, this also depends on the discretization of the Dirac matrix
and on the $\Gamma$- and derivative-structure of the creation operators.
Hence predicting the
efficiency of a specific scheme is difficult.
An object frequently appearing in this work is the pseudoscalar loop
$\mathrm{Tr}(\gamma_5M^{-1})$. In our spinor representation
$\gamma_5$ is diagonal so that the na\"{\i}ve picture presented above
may apply. For other, non-diagonal $\Gamma$-structures different
SSP schemes may be more effective. The picture becomes further obscured
since within all the
partitioning schemes there will be residual couplings
between different colour components on the same site too.
Fortunately, these terms are suppressed by the fact that
$S(y|y)$ will be quite colour-diagonal, in particular in the
heavy quark limit. We also investigate combinations of
(S)SP and colour partitioning.

\subsubsection{Hopping parameter expansion}
We have seen above that
stochastic noise components that are
close to the diagonal of the inverse Fermionic matrix
$M^{-1}$ are accompanied
by larger amplitudes than terms that are far off the
diagonal. Therefore the cancellation of near-diagonal noise
requires a comparatively larger number of estimates.
The HPE noise subtraction~\cite{Thron:1997iy} is based on the observation that
one can expand, see Eq.~(\ref{eq:mwilson}),
\begin{equation}
M^{-1}=2\kappa\sum_{i=0}^{\infty}(\kappa D)^i=
2\kappa\sum_{i=0}^{k-1}(\kappa D)^i+(\kappa D)^kM^{-1}\,,
\label{eq:hpe2}
\end{equation}
where $k\geq 1$. For distances between source and
sink that are composed of more than $k$ links,
the first term on the right hand side does not
contribute since $D$ only connects nearest
spacetime neighbours. Therefore, $M^{-1}_{xy}=
[(\kappa D)^kM^{-1}]_{xy}$ for sufficiently large source and
sink separations. However their estimates will differ, $M_{E,xy}^{-1}\neq
[(\kappa D)^kM^{-1}_E]_{xy}$.
The variance of the latter
estimate of $M^{-1}_{xy}$ will be reduced since less noise terms
contribute and in particular the dominant sources
of noise have been removed.
This was for instance exploited
in Refs.~\cite{Bali:2005pb,Bali:2005fu}.

\begin{figure}
\includegraphics[width=.48\textwidth,clip]{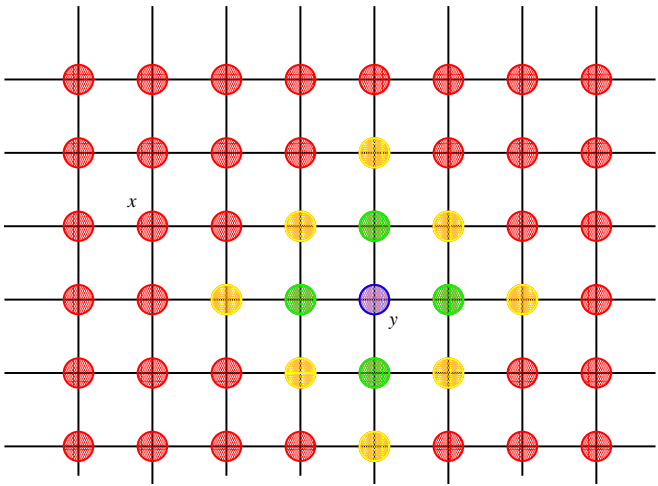}
 \caption{Two dimensional sketch of the effect of HPE.
$y$ indicates the sink and $x$ an arbitrary source site.
One application of $\kappa D$ reduces the blue, two applications
the green and three applications the yellow contributions to the
noise. 
\label{hpa}}
\end{figure}

We illustrate the HPE technique in Fig.~\ref{hpa}:
one application of $\kappa D$ reduces the blue contributions,
two applications the green ones, three applications the yellow ones and so
forth. However, note that unlike in the case of partitioning,
these are not completely removed since they
can propagate along a longer path to reach the sink, weakening their effect.
It is self-evident that the HPE will be particularly effective
for heavy quarks.

Here we also study closed loops, i.e.\ $x=y$.
Obviously, only even powers of $D$ contribute to
$\mathrm{Tr}\,(M^{-1}\Gamma)$, where $\Gamma\in\{{\mathbb 1},
\gamma_{\mu},\sigma_{\mu\nu},\gamma_{\mu}\gamma_5,\gamma_5\}$.
We can write,
$\mathrm{Tr}\,(M^{-1}\Gamma)=\kappa^k\mathrm{Tr}\,(D^kM^{-1}\Gamma)$
for $k\leq k_{\max}$. For $\Gamma=\gamma_5$ and $\Gamma=\gamma_i$
for the clover action, $k_{\max}=2$.
Moreover, the lowest non-vanishing terms have been calculated
analytically and can be computed and corrected for
exactly (unbiased noise
subtraction)~\cite{Thron:1997iy,Wilcox:1999tc,Mathur:2002sf,Deka:2008xr}.
Here we do not attempt to do this but we restrict ourselves to
$k=2$ instead.

The HPE comes with very little computational overhead and,
unlike in the case of partitioning, no additional solves
are required.

\subsubsection{Recursive noise subtraction}
Within the RNS method~\cite{Ehmann:2009ki}
the off-diagonal terms of
Eq.~(\ref{eq:reduce}) are estimated and subsequently
corrected for,
\begin{align}
 M^{-1} & = \overline{|s\rangle \langle \eta|}  +  M^{-1} \left({\mathbb 1}-\overline{|\eta \rangle \langle \eta|}\right) \nonumber \\
 & \approx  \overline{|s\rangle \langle \eta|}  + 
 \overline{|s\rangle \langle \eta|} \left({\mathbb 1}-\overline{|\eta \rangle \langle \eta|}\right)\,,
\label{eq:rns}
\end{align}
in the hope to arrive at an improved estimate.
The second term of the last line of the above equation
involves additional inner products
$\langle \eta_i|\eta_j\rangle$. For $i\neq j$ these fluctuate
randomly but we sum over $12V\gg N^2$ such terms,
where $N$ is the number
of estimates.
Clearly, the procedure can only work if this inner product
is taken over a smaller subspace. Therefore, we compute the random matrix
$\overline{|\eta\rangle\langle\eta|}$ only in the spin-colour subspace,
setting all elements connecting different sites to zero.

\begin{figure}
\includegraphics[height=.48\textwidth,clip,angle=270]{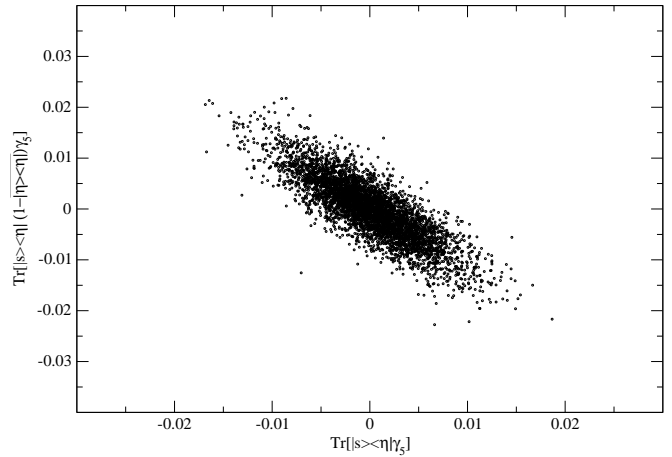}
 \caption{Scatter plot for the RNS, see Eq.~(\protect\ref{eq:rns})
for the pseudoscalar loop.
\label{rns}}
\end{figure}

In Fig.~\ref{rns} we display the correlation between the
two terms of Eq.~(\ref{eq:rns}) for the pseudoscalar loop
$\mathrm{Tr}(M^{-1}\gamma_5)$. Since the two quantities are
anti-correlated, adding them together reduces the statistical
error. So far we have assumed the coefficient of the second
term of Eq.~(\ref{eq:rns})
to be unity. However, realizing that this term
is an unbiased estimate of zero, one can generalize this method,
e.g.\ by allowing for an adjustable coefficient.
Moreover, different terms involving powers
of ${\mathbb 1}-\overline{|\eta\rangle\langle\eta|}$
and/or a different subspace where this matrix assumes
non-trivial values could be implemented. However, we have not
explored these possibilities any further.

\subsubsection{Comparison of methods}
We will apply the methods presented so far, to calculate
disconnected quark loop contributions to the charmonium
spectroscopy. An important example is the correlation of
two zero-momentum projected disconnected loops, e.g.,
\begin{equation}
\label{eq:pion2}
\sum_{x,{\mathbf y}}
\left\langle
\mathrm{Tr}\left[(\Phi M^{-1}\Phi)_{yy}\gamma_5\right]
\mathrm{Tr}
\left[(\Phi M^{-1}\Phi)_{xx}\gamma_5\right]\right\rangle\,,
\end{equation}
for the $\eta_c$ mass where $y_4=x_4+t$. $\Phi=\Phi^{\dagger}$
denotes a Wuppertal smearing function
and the traces are over spin and colour. Note that the above 
traces are real, due to $M^{\dagger}=\gamma_5M\gamma_5$. 
We also remark that we used two separate sets
of noise vectors to estimate the two traces, as one has to do.
To compare the methods we choose $t=a$ and $n_{\mathrm{wup}}=10$ on
ensemble \mycirc{1}.

\begin{table}
\caption{Gain factors of the noise reduction methods tested
for the pseudoscalar disconnected correlator
Eq.~(\protect\ref{eq:pion2}) at the charm quark mass. $k$
denotes the number of $\kappa D$ applications.
\label{overview_tab}}
\begin{center}
\begin{ruledtabular}
\begin{tabular}{ccc}
k & 0 & 2\\\hline
no partitioning& 1 &2.89\\
spin & 1.43&6.32\\
colour&1.80&5.06\\
spin + colour&2.52&10.24\\
odSSP&2.30&5.42\\
obdSSP&1.97&7.16\\
obdSSP + colour&3.63&11.80\\
RNS&1.87&5.44
\end{tabular}
\end{ruledtabular}
\end{center}
\end{table}

The improvements in terms of the real computer time spent to achieve
the same stochastic error are displayed in Table~\ref{overview_tab}
for different combinations of methods.
All overheads are included, except for the (negligible) cost of
the two $\kappa D$ applications. $k$ denotes the power of $\kappa D$
applied to the solution vector. The biggest net gain factor
amounts to almost 12.
Based on these numbers we decide to use the obdSSP and colour partitioning,
together with the HPE for this type of diagram.

\section{The Spectrum}
\label{sec:spectrum}
In this section we calculate the spectrum created by $\bar{c}c$ quark
bilinears, neglecting charm quark annihilation and light quark
creation diagrams.
We first discuss our operator basis and then
the spectroscopy results. The variational method also reveals
information about the spatial structure of the underlying
states. We will discuss this as well as the mixing between
$S$- and $D$-wave operators in the $J^{PC}=1^{--}$ vector channel.

\begin{table}
\caption{Quark bilinears that we use ($s_{ijk}=|\epsilon_{ijk}|$,
also see Eq.~(\protect\ref{eq:defD})).
\label{operators}}
\begin{center}
\begin{ruledtabular}
\begin{tabular}{ccccc}
 name & $\mathrm{O_h}$ repr. & $J^{PC}$ & state & operator  \\
 \hline
 $\pi$ & $A_1$ & $0^{-+}$ & $\eta_c$ & $\gamma_5$   \\
 $\rho$ & $T_1$ & $1^{--}$ & $J/\Psi$ & $\gamma_i$  \\
 $b_1$ & $T_1$ & $1^{+-}$ & $h_c$ & $\gamma_i\gamma_j$  \\
 $a_0$ & $A_1$ & $0^{++}$ & $\chi_{c0}$ & 1   \\
 $a_1$ & $T_1$ & $1^{++}$ & $\chi_{c1}$ & $\gamma_5\gamma_i$  \\
 $(\rho\times\nabla)_{{T_2}}$ & $T_2$ & $2^{++}$ & $\chi_{c2}$ & $s_{ijk}\gamma_j\nabla_k$  \\
 $(\pi\times D)_{{T_2}}$ & $T_2$ & $2^{-+}$ &  & $\gamma_4\gamma_5 D_i$  \\
 $(a_1\times\nabla)_{{T_2}}$ & $T_2$ & $2^{--}$ &  & $\gamma_5 s_{ijk}\gamma_j\nabla_k$  \\
 $(\rho\times D)_{{A_2}}$ & $A_2$ & $3^{--}$ &  & $\gamma_i D_i$  \\
 $(b_1\times D)_{{A_2}}$ & $A_2$ & $3^{+-}$ &  & $\gamma_4\gamma_5\gamma_i D_i$  \\ 
 $(a_1\times D)_{{A_2}}$ & $A_2$ & $3^{++}$ &  & $\gamma_5 \gamma_i D_i$  \\
 $(a_1\times B)_{{T_2}}$ & $T_2$ & $2^{+-}$ & {\scriptsize exotic} & $\gamma_5 s_{ijk}\gamma_j B_k$  \\ 
 $(b_1\times\nabla)_{{T_1}}$ & $T_1$ & $1^{-+}$ & {\scriptsize exotic} & $\gamma_4\gamma_5\epsilon_{ijk}\gamma_j\nabla_k$
\end{tabular}
\end{ruledtabular}
\end{center}
\end{table}

\subsection{Extraction of masses}
The operators that we employ to create the charmonium states
are based on Ref.~\cite{Liao:2002rj}.
We restrict ourselves to the subset of these operators for which
we are able to obtain meaningful signals. These
are displayed in Table~\ref{operators}, together
with their irreducible lattice representations and the
corresponding lowest continuum spin assignments, see Table~\ref{tab:lat1}.
These assignments are of course not unique. For instance a
radial $T_1$ excitation can, in the continuum limit,
very well correspond to $J=3$ or $J=4$, see Table~\ref{tab:lat2}.
We label the operators by the names of the corresponding isovector
mesons (which in nature are no charge eigenstates) since
we are most familiar with these.

Note that in the non-relativistic quark model,
\begin{equation}
P=(-)^{L+1}\,,\quad C=(-)^{L+S}\,,
\end{equation}
where $S\in\{0,1\}$ and $J\in\{L+S,L,|L-S|\}$.
The states that cannot be accommodated in this way,
namely $0^{--}, 0^{+-}, 1^{-+}, 2^{+-}, 3^{-+}, \ldots$
are commonly referred to as ``spin-exotic'' states.
As one can see from the $1^{-+}$ example of Table~\ref{operators}
these exotic states do not need to be tetraquarks/molecules or hybrid mesons.
With relativistic quarks, local bilinears are not restricted to
$0^{-+}$ and $1^{--}$ anymore but e.g.\ $1^{+-}$ can be created with
$L=0$. In this case the ``exotic'' $1^{-+}$ state
is merely a $1^{+-}$ quark bilinear in a
$P$-wave. We also remark that in QCD with finite quark masses
$L$ is not a good quantum number. However, it may still provide us
with some guidance if the quarks are heavy. We will address this
issue in Sec.~\ref{Se:vecmix} below.

In Table~\ref{operators}, $\nabla$ represents a
covariant spatial derivative and $D$ and $B$ the symmetrized and
anti-symmetrized combinations,
\begin{equation}
\label{eq:defD}
D_i = s_{ijk}\nabla_j \nabla_k\,,\quad
B_i = \epsilon_{ijk}\nabla_j \nabla_k\,,
\end{equation}
with $s_{ijk}=|\epsilon_{ijk}|$ and we sum over $j$ and $k$.
All operators containing a covariant derivative implicitly also
include gluonic contributions but then any $P$- or $D$-wave
will include derivatives and one would hardly call these
hybrid mesons.
However, the $B_i$-operators not only contain
the vector potential but they are proportional to components
of the field strength tensor itself. This
is as close to a valence gluon
as one can get. A strong coupling of a physical state to
this operator may then indicate a large hybrid meson content.
The $1^{-+}$ charmonium is considered a prime hybrid candidate. 
However, we find the operators $\epsilon_{ijk}\gamma_jB_k$
($T_1$ representation, not listed in the Table) to be very noisy,
with no compelling evidence of a coupling to the ground state
created by $b_1\times\nabla$. 

For each operator listed in the Table we construct a three by three
cross-correlator matrix, see Sec.~\ref{Se:varMeth},
with different smearing levels, see Sec.~\ref{sec:fermsmear}.
We apply the same smearing to quark and antiquark.
The smearing parameters have been optimized for several states
as described in Sec.~\ref{sec:fermsmear},
so that point-smeared effective masses are relatively constant
for the narrow operator and approach their asymptotic values from
below for the wide operator.
We only consider the two lowest lying masses reliable
since the highest state contained in the basis may be
polluted by even higher excitations.
In Ref.~\cite{Dudek:2007wv} a similar approach was used
to calculate the spectra of excited states in the quenched approximation.

\begin{figure*}\begin{center}
\includegraphics[width=.9\textwidth,clip]{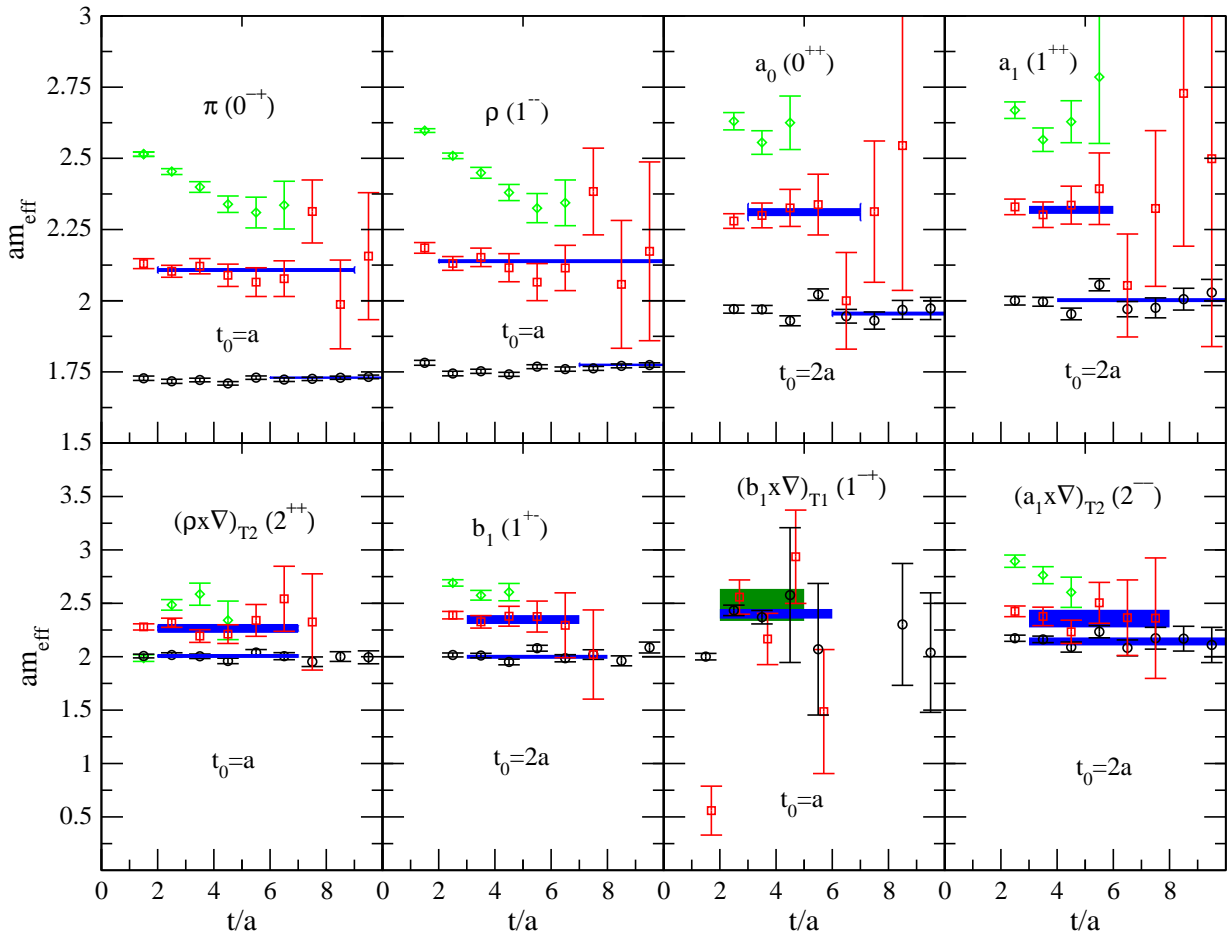}
\end{center}
\caption{Effective masses for some operators of Table~\protect\ref{operators}
obtained on ensemble \mycirc{1}.
The fit ranges and errors are indicated by horizontal lines.
The $t_0$ values (in lattice units) refer to the respective
normalization times Eq.~(\protect\ref{evp}).
\label{effmass}}
\end{figure*}

\begingroup
\squeezetable
\begin{table}
\caption{Fitted masses obtained on ensemble \mycirc{1} for the first
two eigenvalues in each channel. The normalization time $t_0$
and the corresponding fit ranges are also given. The errors are
only statistical and we give the lowest continuum $J^{PC}$ from which
the lattice representation can be subduced.}
\label{tab:spectrum}
\begin{ruledtabular}
\begin{tabular}{ccccccc}
 operator &  $J^{PC}$ & $t_0/a$ & $m_{1}/$MeV & range & $m_{2}$/MeV & range  \\
 \hline
 $\pi$ & $0^{-+}$ & 1 & 2993~~(4) & 5--12 & 3645~(19) & 1--8  \\
 $\rho$ & $1^{--}$ & 1 & 3070~~(6) & 7--12 & 3699~(24) & 1--7  \\  
 $b_1$ & $1^{+-}$ & 2 & 3457~(22) & 2--7 & 4060~(65) & 1--5  \\
 $a_0$ & $0^{++}$ & 2 & 3381~(19) & 4--12 & 3996~(48) & 1--5  \\
 $a_1$ & $1^{++}$ & 2 & 3462~(20) & 3--11 & 4011~(52) & 1--5  \\
 $(\rho\times\nabla)_{T_2}$ & $2^{++}$ & 1 & 3471~(19) & 1--6 & 3917~(46) & 1--6 \\
 $(\pi\times D)_{T_2}$ & $2^{-+}$ & 1 & 3756~(32) & 1--9 & 3995(141) & 1--6 \\
 $(a_1\times\nabla)_{T_2}$ & $2^{--}$ & 2 & 3706~(27) & 1--10 & 4076~(83) & 1--6 \\
 $(\rho\times D)_{A_2}$ & $3^{--}$ & 1 & 3782~(35) & 1--8 & 4815~(92) & 1--6 \\
 $(b1\times D)_{A_2}$ & $3^{+-}$ & 1 & 3995~(50) & 2--6 & 5365~(76) & 1--3 \\
 $(a_1\times D)_{A_2}$ & $3^{++}$ & 2 & 3993~(54) & 1--5 & 5008(287) & 1--4 \\
 $(b_1\times\nabla)_{T_1}$ & $1^{-+}$ & 1 & 4154~(54) & 1--5 & 4297(181) & 1--4 \\
 $(a1\times B)_{T_2}$ & $2^{+-}$ & 1 & 4614(220) & 1--9 & 4643(254) & 1--8
\end{tabular}
\end{ruledtabular}
\end{table}
\endgroup
\subsection{Discussion of the results}
\label{sec:discuss}
We display effective masses 
for ensemble \mycirc{1} (see Table~\ref{confdetail_tab}),
obtained after
diagonalizing the correlation matrices at the normalization
time $t_0$, see Eq.~(\ref{evp}), for some of the channels
in Fig.~\ref{effmass}.
Only the lowest two masses are fitted and the fit ranges are
indicated by the blue lines.
The extracted mass values, together with these fit ranges, are
displayed in Table~\ref{tab:spectrum}.
In this table we also assign the lowest possible continuum
spin to each channel. Note however, that radial excitations
of both $J=1$ ($=T_1$) and $J=2$ ($=T_2$) states could in principle
correspond to $J=3$. In particular, this possibility cannot
be excluded for the excitations of the $T_1$ states
that we have labelled as $1^{+-}$ and $1^{++}$ and for the
$T_2$ state labelled as $2^{++}$.

The ground states and first excitations of the standard $S$- and $P$-wave
states $\eta_c$, $J/\Psi$, $h_c$ and $\chi_{cJ}$
exhibit good signals and stable plateaus. The effective
masses of higher spin states are naturally noisier
and thus complicate the fits.

A particularly interesting channel is the $1^{-+}$. Although this is
a spin-exotic state it can be well accessed by the
$b_1\times\nabla$ operator that does not contain
an explicit chromo-magnetic field. However, $\nabla$
contains a link variable and may allow for a gluonic excitation.
The quality of the effective masses arising from the
$b_1\times\nabla$ operator is not high
but the fit yields reasonable errors.
The two lowest lying states are very close. In fact, within their errors
the effective masses are overlapping so that in the statistical
analysis it was necessary to sort the jackknifes according to the
proximity of the eigenvectors to the ones obtained on the original ensemble.
This may hint at a hybrid nature of this channel. Static hybrid potentials
are repulsive at short distances so that, within a potential model, we
may expect smaller energy gaps between radial
excitations~\cite{Perantonis:1990dy,Collins:1997cb,Juge:1999ie}.

\begin{figure*}
\begin{center}
\includegraphics[width=.8\textwidth,clip]{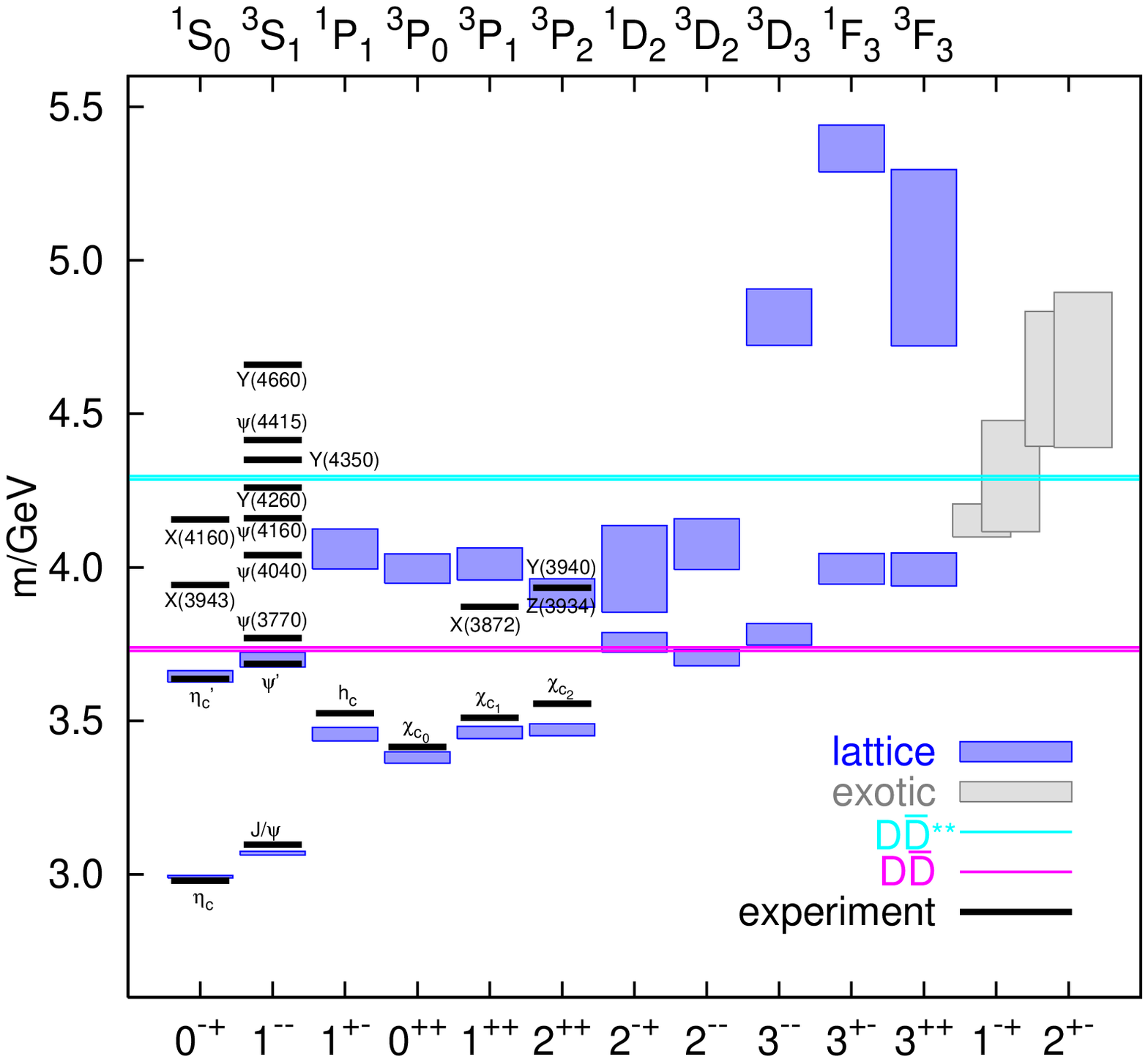}
\end{center}
 \caption{The predicted spectrum, together with the experimental values on ensemble \mycirc{1}, see Table~\protect\ref{confdetail_tab}.
\label{spectrum}}
\end{figure*}

The computed spectrum is displayed in Fig.~\ref{spectrum},
together with the experimental values\footnote{In some
cases their $J^{PC}$ assignment is still under debate. For instance
for the $X(3872)$ that we list as a $1^{++}$ state a $2^{-+}$ assignment
is not ruled out.} 
 We have used the spin-averaged
$1\overline{S}$ mass Eq.~(\ref{eq:sa}) to fix the charm quark mass.
The other states are predictions.
Note, however, that we underestimated $m_{1\overline{S}}$ by about 15~MeV.
The main effect of this is that all the predictions should be shifted up
by 15~MeV.
Keeping this in mind we observe all spin-averaged states below
threshold coming out in qualitative agreement with the experimental
data~\cite{Nakamura:2010zzi}.
We indicate the experimental
$D\overline{D}$ and $D\overline{D}^{**}$ open charm thresholds
as horizontal lines. Negative parity states cannot decay
into $D\overline{D}$ and the $D\overline{D}^{**}$ threshold
is believed to play a more prominent r\^ole in the decay of hybrid
mesons than $D\overline{D}^{*}$.

The spin-averaged $1\overline{P}$-$1\overline{S}$
splitting is underestimated, relative to experiment while
the $2\overline{S}$-$1\overline{S}$ splitting comes out
right. There may be issues with the scale setting, due to the
unrealistic sea quark content. We have also remarked in
Sec.~\ref{sec:gauge} above that there are reasons
to believe~\cite{najjar}
that the lattice spacing should be set to $a^{-1}\approx 1.81$~GeV
rather than to the $a^{-1}\approx 1.72$ GeV that we used. This change
would bring the $1\overline{P}$-$1\overline{S}$ splitting
in line with experiment
but result in an overestimated $2\overline{S}$ mass.
This in turn could then be due to a combination of
finite size effects and interferences
with the $D\overline{D}$ threshold in nature that
do not occur on our ensemble, due to the heavy sea quark mass.

From pNRQCD one would, to leading non-trivial order
in $1/m_c$, expect the $S$-wave
finestructure to be determined by the matrix
element~\cite{Bali:2000gf,Pineda:2000sz},
\begin{equation}
\label{eq:nrqcd}
\frac{c_F^2}{3m_c^2}\langle\Psi|V_4(r)|\Psi\rangle\,,
\end{equation}
where to leading order in perturbative QCD,
\begin{equation}
V_4(r)=\frac{32\pi}{3} \alpha_s\delta^3(r)\,.
\end{equation}
$\Psi$ is the non-relativistic charmonium wavefunction,
$\alpha_s$ the strong coupling parameter and
$c_F=1+{\mathcal O}(\alpha_s)$
is a matching coefficient that has only 
recently been
determined in lattice schemes~\cite{Hammant:2011bt}.

This illustrates the very short-distance nature
of the $S$-wave finestructure
that should be affected significantly both
by lattice spacing effects and by differences in the
running of the coupling depending on the sea quark content.
It is also very sensitive to the charm
quark mass. Reducing this
by 5~\% would increase the splitting by 10~\%.
So it is not surprising that we underestimate the
experimental number of 117~MeV
for the $1S$ finestructure splitting. We obtain
$\Delta m_{1S} = m_{J/\Psi}-m_{\eta_c}=77(2)$~MeV on ensemble \mycirc{1},
$\Delta m_{1S} = 88(4)$~MeV on ensemble \mycirc{2}
and $\Delta m_{1S}=130(9)$~MeV on ensemble \mycirc{3}. Indeed, with
lighter sea quark masses this value increases.

For the $2S$ finestructure splitting we obtain
$\Delta m_{2S}=m_{\Psi(2S)}-m_{\eta_c'}=
54(6)$~MeV on ensemble \mycirc{1} and
$\Delta m_{2S}=56(8)$~MeV on ensemble\footnote{
On ensemble \mycirc{3} where the radial excitations
are seriously affected by the finite volume
we get $\Delta m_{2S}=177(66)$~MeV.}
\mycirc{2}, in agreement
with the experimental value of 49(4)~MeV.
In view of the disagreement of the $1S$ splitting
this is quite
surprising since
one would have expected a lot of the systematics to
cancel from the ratio of the $2S$ hyperfine splitting
over the $1S$ splitting, see Eq.~(\ref{eq:nrqcd}).
We may therefore wonder whether either the physical
$\eta_c$ or the $\Psi(2S)$ states are unusually low, 
due to contributions from  quark line disconnected diagrams.
In the first case our neglection of $\bar{c}c$ annihilation
diagrams may be relevant while in the second case omitting
$\bar{q}q$ creation (and the use of
unphysically heavy light quark masses)
would be the dominant effect(s), see Secs.~\ref{sec:etamix}
and \ref{sec:moleculemix} below, respectively.

We remark that we also underestimate the $P$-wave
finestructure. This is expected too and again mostly
due to lattice spacing effects and an unrealistic
sea quark content.
We also notice that in our approximation where the open charm thresholds are
much higher than in nature the $Z(3934)$ (recently renamed into
$\chi_{c2}(2P)$~\cite{Nakamura:2010zzi}) may indeed be
associated with the $\chi_{c2}'$ state while the $X(3872)$
certainly is lighter than one would have expected from an excited $P$-wave.
However, in the first case we cannot exclude the possibility
that we have misidentified a $3^{++}$ state as $2^{++}$, in particular
since this comes out lighter than the other two $\chi_c'$ multiplet masses.
Finally, the proximity of the two $1^{-+}$ states as well as of
the two $2^{+-}$ states
may indicate a hybrid nature of these spin-exotic charmonia. 
We have not detected such indications in any of the other channels.
With the exception of the $A_2$ ($J=3$),
in these cases the radial excitations
are lower in energy than these spin-exotic states.

\subsection{``Wavefunctions''}
\label{Se:wavFunc}
In Sec.~\ref{Se:varMeth} we have introduced couplings between
an operator $\hat{O}_i$, $i=1,\ldots, N$,  and a physical state $|n\rangle$,
$v_i^n=\langle 0|\hat{O}_i|n\rangle$. These will be approximated,
up to a rotation and normalization, see below, by the $\psi^n(t,t_0)$ of
Eq.~(\ref{evp}). We employ the normalization, $\sum_i|\psi^n_i(t,t_0)|^2=1$.
In the pseudoscalar channel our operators read,
\begin{equation}
\label{eq:opdef}
\hat{O}_i=\sum_{{\mathbf x},{\mathbf y}}
\bar{c}({\mathbf y})\Phi_i({\mathbf y}-{\mathbf x})\gamma_5 c({\mathbf x})\,.
\end{equation}
Here $\Phi_i$ denotes the square of the Wuppertal smearing operator,
Eq.~(\ref{wupp}), since we apply this to quark and antiquark
fields. $\Phi_i$ is translationally invariant and will only depend on
the difference ${\mathbf y}-{\mathbf x}$.

We employ a $N=4$ dimensional trial basis consisting of
$n_{\mathrm{wup}}=0, 5, 10$ and 40 iterations. This means that
the $\Phi_i$ contain twice these numbers of iterations.
Folding these smearing functions
with the asymptotic
eigenvectors results in a new smearing function,
\begin{equation}
\Psi^n({\mathbf x})=\sum_i\psi^n_i\tilde{\Phi}_i({\mathbf x}) \,,
\end{equation}
where
\begin{equation}
\tilde{\Phi}_i=\frac{1}{d_i}\sum_j\Phi_jU_{ji}\,.
\end{equation}
$U\in\mathrm{SO}(N)$ diagonalizes $C(t_0)$ and
$d_j>0$
are the square roots of its eigenvalues, 
\begin{equation}
\label{eq:rotate}
\left[C^{-1/2}(t_0)C(t)C^{-1/2}(t_0)\right]_{ij}=
\frac{\left[U^{\dagger}C(t)U\right]_{ij}}{d_id_j}\,,
\end{equation}
see Eq.~(\ref{evp}). In particular this means
that the operators constructed from
$\tilde{\Phi}_j$, $\hat{\tilde{O}}_j$ [see Eq.~(\ref{eq:opdef})]
create states that are orthonormal at\footnote{
If corrections from truncating the basis
can be neglected for $t_0=0$ (which is unlikely) then,
at this $t_0$, $\psi_i^n\rightarrow
v_i^n$ at large times $t$.}
$t= t_0$:
$\langle \tilde{O}_i(t_0)\tilde{O}^{\dagger}_j(0)\rangle=\delta_{ij}$.
Also note that if we neglect the coupling of the operator $\hat{O}_i$ to
states with energies bigger than $E_N$, $d_i\propto \exp(-E_it_0/2)$.

If we had perfect overlap with the respective physical states
then we could choose $t_0=t=0$. In this case, in the non-relativistic
limit, where we do not encounter particle-antiparticle creation,
one may identify
$|\Psi^{n\dagger}({\mathbf x})\Psi^n({\mathbf x})|$
with the quantum mechanical
probability density.
On a qualitative level, we may still think of
$\Psi^n({\mathbf x})$ as the wavefunction of the $n$th
state.
The used ensemble \mycirc{1} is unfortunately too coarse
to resolve the node structure of the gauge invariant $|\Psi^{n\dagger}\Psi^n|$.
However, one can also plot a diagonal colour component of
$\Psi^n({\mathbf x})$, after fixing to Coulomb gauge. In fact
the APE smeared gauge links are so close to unity that
it is hard to resolve the differences between Coulomb gauge
fixing and a non-gauge fixed component from a plot. 

\begin{figure}
\begin{center}
\includegraphics[width=.38\textwidth,clip]{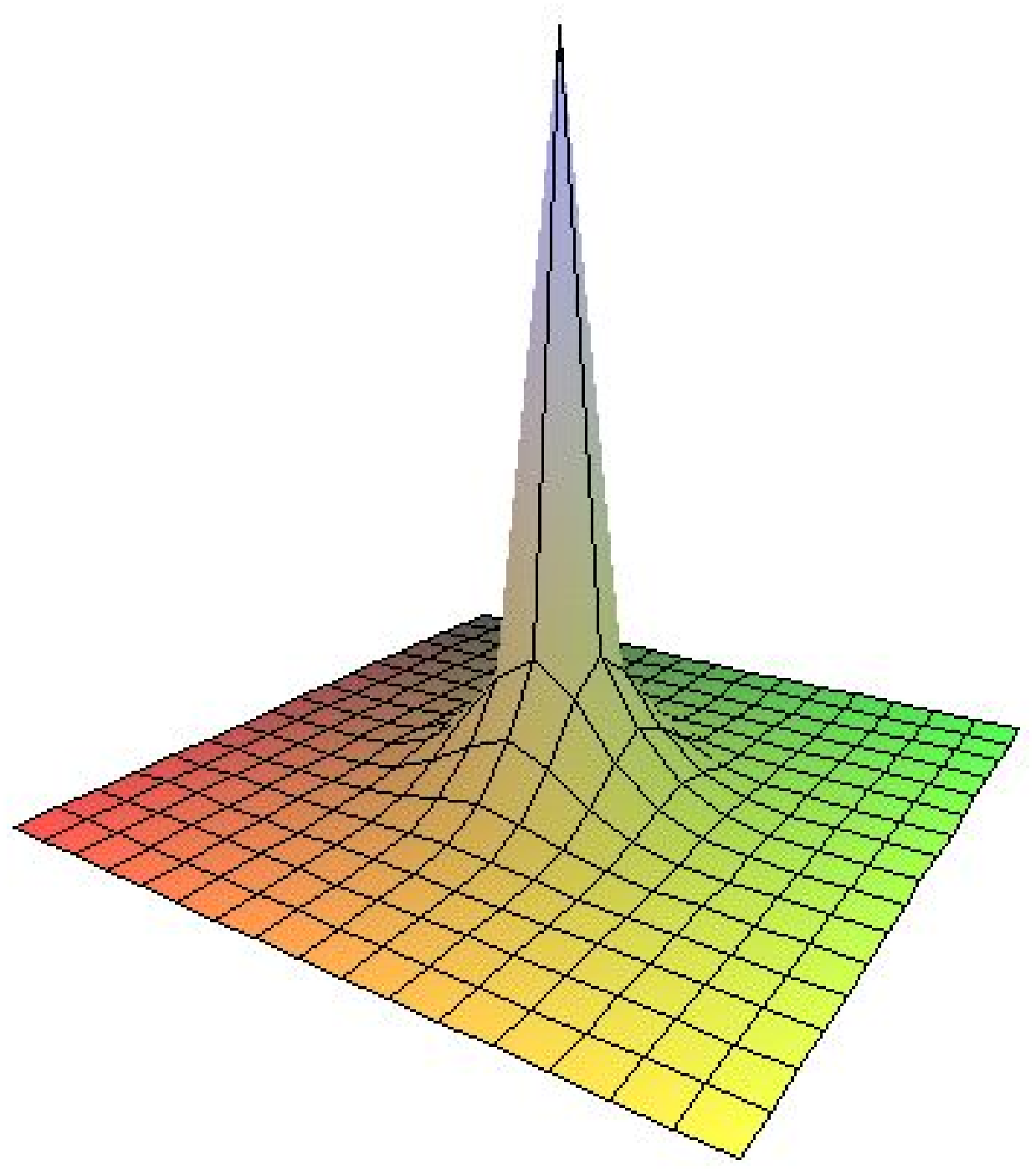}\\
\includegraphics[width=.38\textwidth,clip]{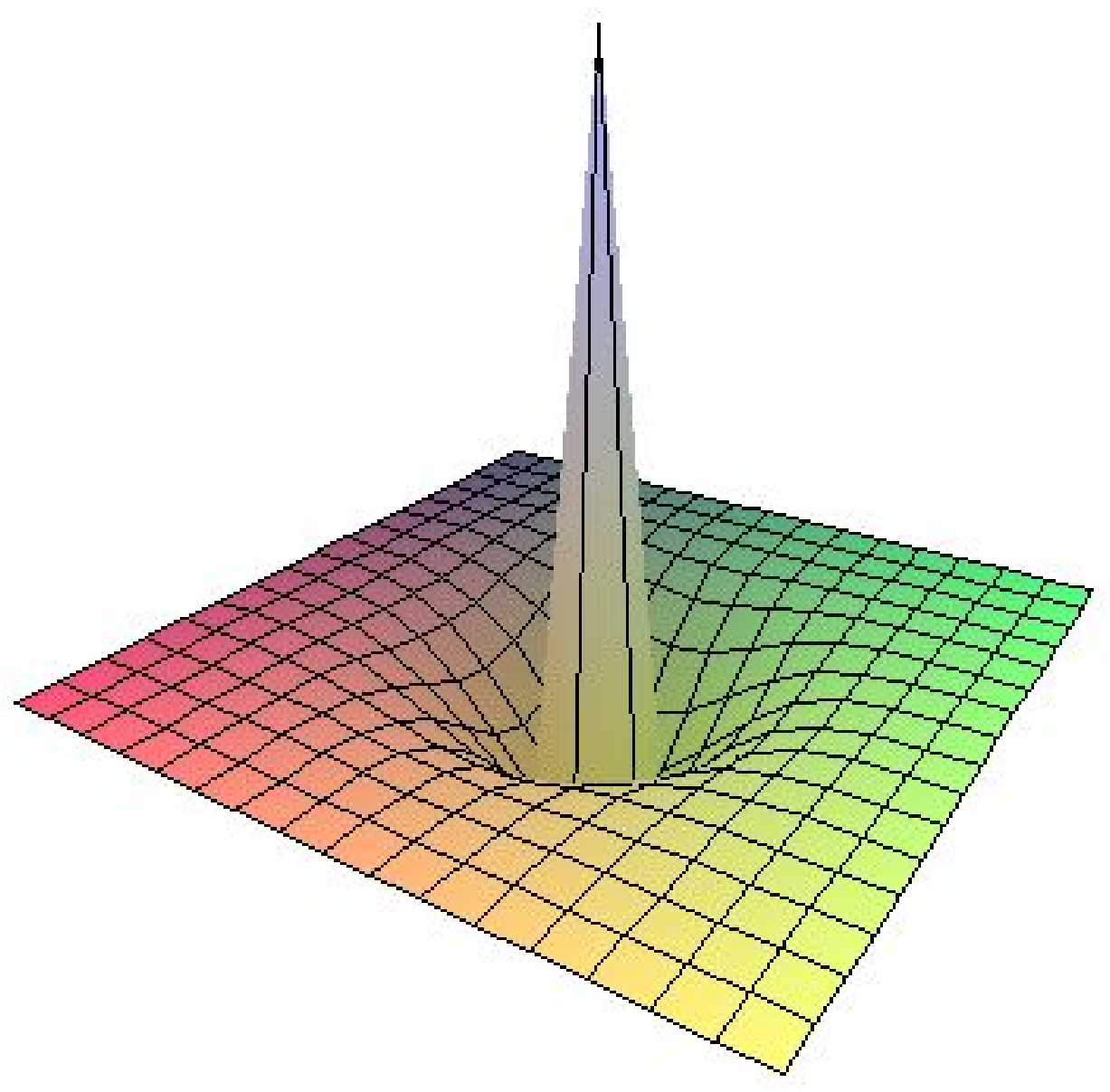}\\
\includegraphics[width=.38\textwidth,clip]{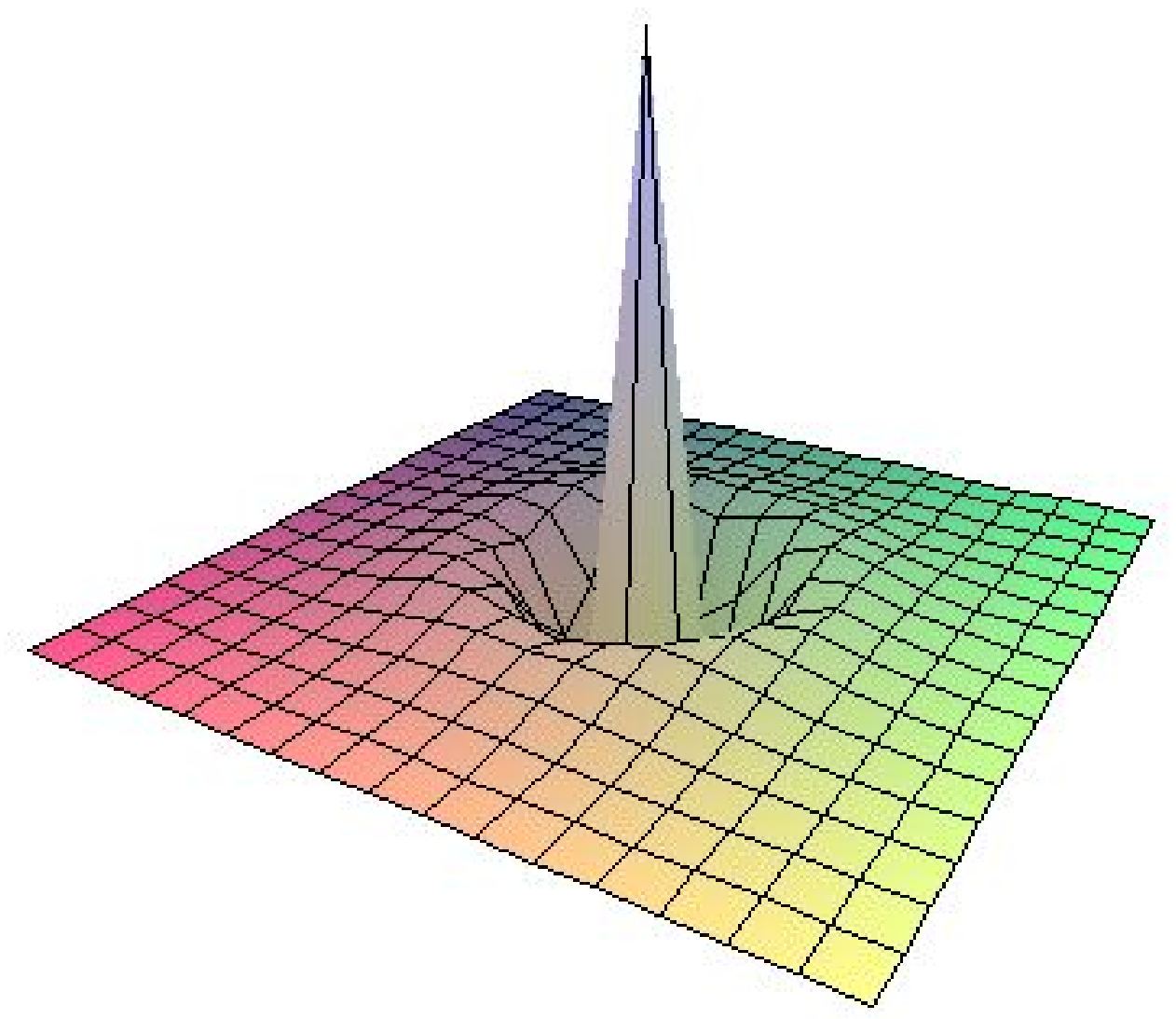}
\end{center}
\caption{The $1S$, $2S$ and $3S$ pseudoscalar ``wavefunctions''.}
\label{wf}
\end{figure}

In Fig.~\ref{wf} we show a two dimensional cross section
of one colour component of the
normalized wavefunctions $\Psi^n({\mathbf x})$, $n=1,2,3$.
In spite of the small basis and lattice volume the node structure
is consistent with the $1S$, $2S$ and $3S$ assignments, with no visible
pollution from higher Fock states or $D$-waves.
For the $1S$ ``wavefunction'' we obtain an rms radius,
$\Delta r=\langle r \rangle_{\mathrm{rms}} = \sqrt{\sum_{V} r^2 |\Psi|^2}
\approx 0.39$~fm. This compares reasonably well with the
infinite volume continuum potential model expectation of
about 0.4~fm \cite{Bali:1998pi}.

\subsection{Mixing in the vector channel}
\label{Se:vecmix}
Due to its direct production in electron-positron annihilation,
the vector channel is rich in experimentally confirmed resonances,
as can be seen in Fig.~\ref{spectrum}. Of great interest is the
inner structure of these states, in particular of the $\Psi(2S)$
and the $\Psi(3770)$ states, which have a mass difference of only
about 90~MeV and both are close to the $D\overline{D}$ open charm
threshold. While $J/\Psi$ is dominated by $1S$ quark-antiquark
configurations, its excitations may exhibit a more complex structure.

As the name suggests, the $\Psi(2S)$ is thought to be a radial excitation.
Since $\Psi(3770)$ is so close in mass, it is very improbable that
it is excited in a further, higher radial vibration mode.
One possibility which we investigate here is an orbital excitation
where the quark-antiquark pair is in a relative
$D$-wave.
We start from an operator basis consisting
of three $S$-wave and two $D$-wave interpolators, 
\begin{align}
(\bar{c}\gamma_ic)_0\,,\quad
(\bar{c}\gamma_ic)_{20}\,,\quad
(\bar{c}\gamma_ic)_{80}\,,\\
(\bar{c}s_{ijk}\gamma_jD_kc)_0\,,\quad
(\bar{c}s_{ijk}\gamma_jD_kc)_{80}\,,\nonumber
\end{align} 
where $D_k$ is defined in Eq.~(\ref{eq:defD}) and
$s_{ijk}=|\epsilon_{ijk}|$.
The subscripts denote the numbers of smearing
iterations, both for the quark and antiquark.
Initially we planned to include hybrid operators like
$\bar{c}\gamma_5B_ic$ into our basis but unfortunately
these provided very poor signals throughout, independent of
the smearing levels.

\begin{figure}
\includegraphics[height=.48\textwidth,clip,angle=270]{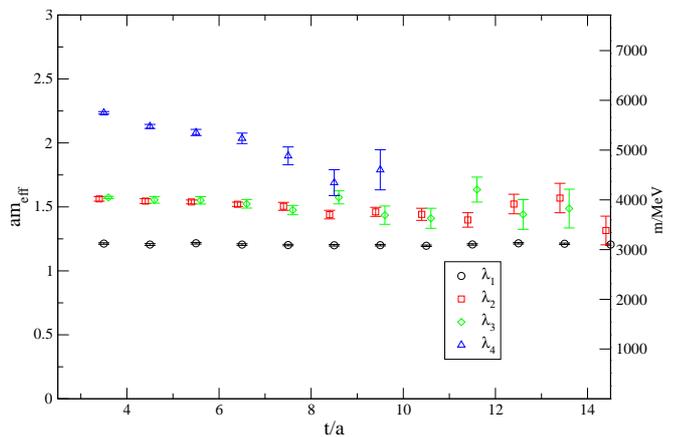}
\caption{Effective masses of the four lowest lying states in the vector channel on ensemble \mycirc{2}.}
\label{vec_ev}
\end{figure} 

\begin{figure}
\includegraphics[width=.48\textwidth,clip]{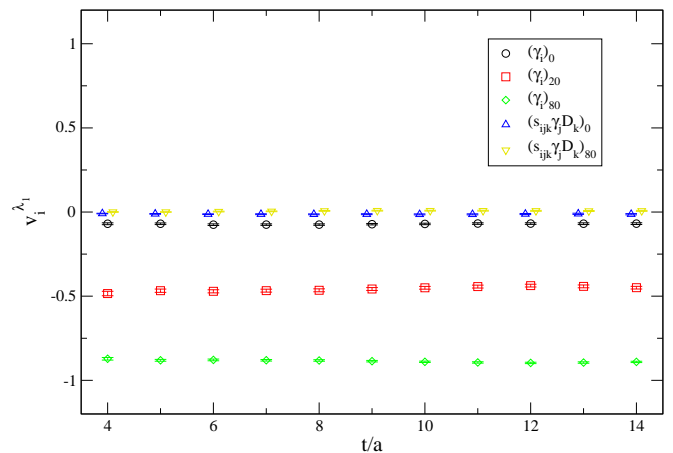}
\caption{Components of the first $1^{--}$ eigenvector,
see Eq.~(\protect\ref{eq:overl}).}
\label{vec_evec1}
\end{figure} 

\begin{figure}
\includegraphics[width=.48\textwidth,clip]{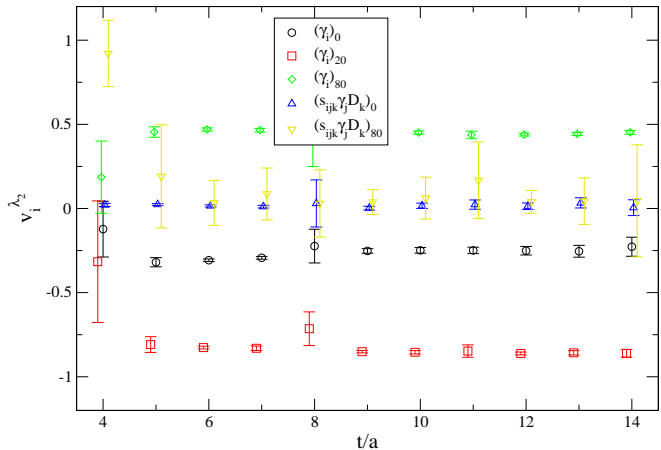}
\caption{Components of the second $1^{--}$ eigenvector.}
\label{vec_evec2}
\end{figure} 

\begin{figure}
\includegraphics[width=.48\textwidth,clip]{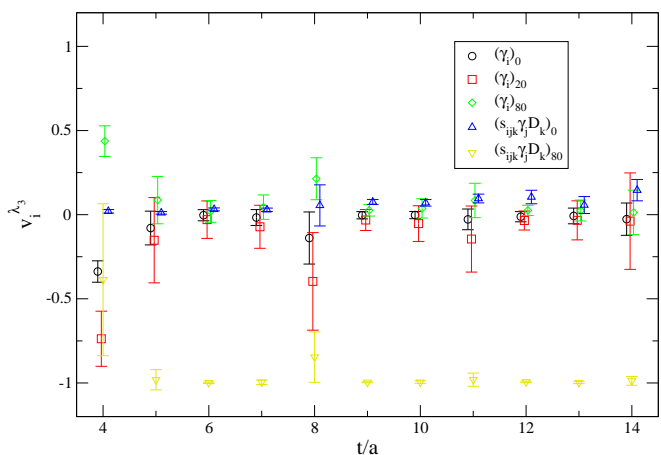}
\caption{Components of the third $1^{--}$ eigenvector.}
\label{vec_evec3}
\end{figure} 

This mixing analysis is performed on ensemble \mycirc{2},
see Table~\ref{confdetail_tab}, with $t_0=3a$.
We display the lowest four effective masses in Fig.~\ref{vec_ev}.
Indeed, the second and third eigenvalues lie very close.
The fourth eigenvalue may be identified with the $\Psi(4040)$.

The eigenvector components reveal
the overlaps between the trial operators and the physical
eigenstates.
The correlation matrix that is real in our case
has the normalization ambiguity
$C_{ij}(t)\mapsto e_ie_jC_{ij}(t)$, where $i=1,\ldots,N$ and $N$
is the dimension
of the operator basis. One convenient choice is
$C_{ii}(0)=1$. The orthogonal transformation $U$
that diagonalizes the correlation matrix at the time $t_0$ and the
eigenvalues at this time $d_i^2$ of
Eq.~(\ref{eq:rotate}) depend on this initial normalization
of $C(t)$.
Following the discussion of Secs.~\ref{Se:varMeth} and \ref{Se:wavFunc},
see Eqs.~(\ref{evp}) and (\ref{eq:rotate}), we can
define effective overlaps,
\begin{equation}
\label{eq:overl}
v_i^n(t,t_0)=\left[\sum_k\left(\frac{\psi_k^n}{d_k}\right)^{\!\!2}\right]^{\!-1/2}\sum_jU_{ij}\frac{\psi_j^n(t,t_0)}{d_j}\,,
\end{equation} 
that in the limit $t_0\gg 0$, $t\gg t_0$ will approach
$v_i^n=\langle 0|\hat{O}_i|n\rangle$, up to an overall factor.
The effective overlaps do not depend on
the normalization choices $e_i$ of $C_{ij}(t)$.
Our normalization
$\sum_i|v_i^n(t,t_0)|^2=1$ also implies
$\sum_n|v_i^n(t,t_0)|^2=1$ which is
equivalent to
ignoring any effects of higher lying states. 
In Figs.~\ref{vec_evec1}, \ref{vec_evec2} and \ref{vec_evec3} we display
the first three $v_i^n(t,t_0=3a)$.

As one would expect the ground state, the $J/\Psi$, does not
receive any contributions
from the two $D$-wave operators. Interestingly, the second
eigenstate, that is energetically very close to the third one
(in fact for most $t$-values we can only differentiate between
these states by tracing their eigenvector components),
does not ``see'' any $D$-wave operators either.
We remark that at the $t$-values where these second and third
energy eigenvalues differ from each other we encounter
more ``mobility'' of the eigenvector components, see
Figs.~\ref{vec_ev}, \ref{vec_evec2} and \ref{vec_evec3}.
Note the relative sign change in the case of
the first excitation between the local/narrow
and the wide operators, resulting in a node of the spatial
wavefunction, similar to the $2S$ state of Fig.~\ref{wf}.
This strongly suggests a $\Psi(2S)$ assignment for this state.
Conversely, the third eigenvalue only couples to the wide smeared
$D$-wave operator, which obviously leaves it as a candidate for
the $\Psi(3770)$. These results compare reasonably well with the
ones of Ref.~\cite{Dudek:2007wv}. We conclude that the charm
quark is sufficiently heavy for $S$- and $D$-waves
to undergo only very mild mixing. So, at least for
charmonia of masses below 3.8~GeV, it is meaningful to label states
by their angular momenta.
However, we have not yet considered the effect of open charm thresholds.
We will address this question in Sec.~\ref{sec:moleculemix} below.

\section{Mixing between the $\boldsymbol{\eta_{\mathbf c}}$ and the light
$\boldsymbol{\eta}$ meson}
\label{sec:etamix}
Charmonia are flavour-singlet states, however, so far we have
neglected the charm quark annihilation diagram that arises
from Wick contracting the correlation function, $\langle[\bar{c}\Gamma c](t)\,
 [\bar{c}\Gamma c]^{\dagger}(0)\rangle$.
The inclusion of quark line disconnected diagrams
may affect charmonium masses.
In particular the proximity of the mass of the $\eta_c$ meson to that
of the pseudoscalar glueball which may propagate as an intermediate
state after $\bar{c}c$ annihilation may have some effect~\cite{Feldmann:1999uf}.
This glueball mass was
consistently determined in simulations of pure gauge theories
on isotropic lattices to be~\cite{Bali:1993fb} $(2630\pm 290)$~MeV
and on anisotropic lattices as\footnote{We converted
the numbers into units of $r_0=0.467$~fm and ignore
the uncertainty in this scale setting.}~\cite{Chen:2005mg} $(2637\pm 26)$~MeV.
We also know that the light quark analogue of the $\eta_c$, the $\eta'$
meson is much heavier than the light octet pseudoscalar mesons,
due to the $\mathrm{U_A}(1)$ anomaly. Naturally,
chiral symmetry will not play a prominent r\^ole for charm quarks.
However, this does not exclude a remnant effect of the vacuum topology
that may lift the $\eta_c$ mass by a few MeV.

First calculations of the disconnected contribution both
with $n_{\mathrm{F}}=2$ sea quarks and in the quenched
approximation are consistent with no mass shift
of the $\eta_c$ mass~\cite{McNeile:2004wu,deForcrand:2004ia}, 
within statistical errors of about 20~MeV.
More recently, in the quenched approximation the $\eta_c$ mass was
estimated to increase by 1--2~MeV~\cite{Levkova:2010ft} due to disconnected
diagrams and, including sea quarks, this effect may become
1--4~MeV~\cite{Levkova:2010ft}.

\begin{figure}
\includegraphics[width=.48\textwidth,clip]{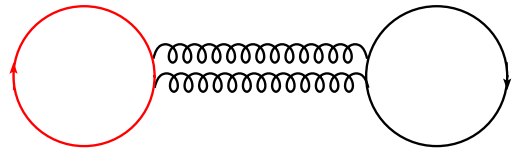}
\caption{The lowest order perturbative QCD graph responsible for
the mixing between $\bar{c}\gamma_5c$ and $\bar{q}\gamma_5q$ states.
The red lines correspond to charm quarks, the black ones to light quarks,
the twiddly ones to gluons.
\label{ccuumix_fig}}
\end{figure}

Naturally, when sea quarks are included 
both physical eigenstates, the $\eta_c$ and the $\eta'$,
will contain light as well as charm valence quarks.
The charm-anticharm component will be dominant within the
$\eta_c$
while the $\eta'$ will almost exclusively contain light quarks.
In our case we employ
$n_{\mathrm{F}}=2$ sea quarks and therefore an isosinglet
$\eta$ state assumes the r\^ole of the $\eta'$.
In addition we have an isovector $\pi$ triplet, instead of the octet.
When the disconnected quark loop is included then, at large Euclidean
times, the $\eta_c$ state will decay to the ground state
in the $J^{PC}=0^{-+}$ channel, which is the $\eta$ meson.
The physical $\eta_c$ will only appear within the tower
of excitations of this ground state. Following
Ref.~\cite{Bali:2005fu} we call this effect ``implicit'' mixing:
the $\bar{c}\gamma_5 c$ state already intrinsically contains
a $\bar{q}\gamma_5 q$ contribution. However, one would expect
the coupling
of the $\bar{c}\gamma_5 c$ creation operator to this state
to be very weak. Otherwise charmonia would not be stable in nature
either. This means that we can treat this as a perturbation.
We decompose the physical Hamiltonian $H=H_0+\lambda H_1+\cdots$,
into a part $H_0$ with\footnote{We omit the $\Gamma$ structure
for convenience.} $\bar{c}c$ and $\bar{q}q$ eigenstates,
without pair creation. The small perturbation
$\lambda H_1$ is then responsible for the mixing. Neglecting
radial, gluonic or multiquark excitations, the physical $\eta_c$
wavefunction of this two-state system reads, to first order
in the small parameter $\lambda$,
\begin{equation}
\label{eq:mix}
|\eta_c\rangle =|\bar{c}c\rangle +
\lambda \frac{\langle \bar{q}q| H_1 | \bar{c}c \rangle}
{E_{\bar{c}c}-E_{\bar{q}q}}|\bar{q}q\rangle\,.
\end{equation}
While we do not know the functional form of $H_1$ or
of the unperturbed wavefunctions, we can evaluate
all the relevant matrix elements on the lattice.
Fig.~\ref{ccuumix_fig} depicts
the graph responsible for this mixing to lowest order in
perturbative QCD.

\begin{figure}[ht]
\includegraphics[width=.48\textwidth,clip]{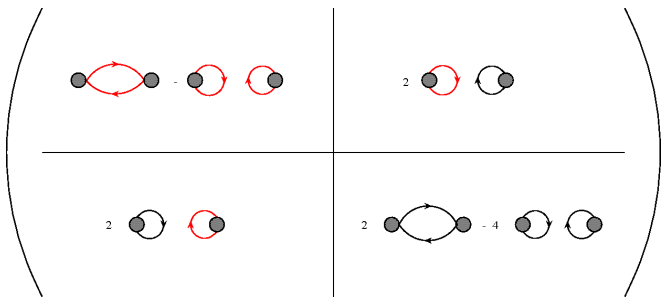}
\caption{Correlation matrix for the mixing between states
created by $\bar{c}\gamma_5 c$
and by $\bar{q}\gamma_5 q$ operators. Red lines represent charm
quarks, black lines light quarks.
\label{mixmatrix_eta}}
\end{figure} 

Obviously, the mixing will depend on the light quark mass value $m_q$.
With decreasing $m_q$ the denominator 
of Eq.~(\ref{eq:mix}) will become larger. However,
we would also expect the matrix element in the numerator to increase
since the probability of creating a light quark-antiquark pair
may increase with decreasing quark mass.
Thus, ideally one would realize several light quark masses
to clarify this issue.

Similar to our investigation of mixing in the vector channel
of Sec.~\ref{Se:vecmix} we also calculate a correlation
matrix here. We choose the basis states,
\begin{align}
(\bar{c}\gamma_5c)_{0}\,,\quad (\bar{c}\gamma_5c)_{10}\,\quad
(\bar{c}\gamma_5c)_{80}\,,\nonumber\\
(\bar{q}\gamma_5q)_{0}\,,\quad (\bar{q}\gamma_5q)_{5}
\,\quad (\bar{q}\gamma_5q)_{40}\,,
\end{align}
where the subscripts denote the number of Wuppertal smearing
iterations. The diagonalization of this matrix at large times
will not only allow us to extract the energy levels but
it will also provide us with qualitative information
on the charm and light quark content of the physical states.
In Fig.~\ref{mixmatrix_eta} we sketch the structure of the mixing matrix
between the $\bar{c}c$ and $\bar{q}q$ sectors, omitting the
different smearing levels. Red lines represent charm quark propagators
and blue lines light quark propagators. The prefactors are due to the
$n_{\mathrm{F}}=2$ mass degenerate sea quark flavours. The upper left
corner contains the $\bar{c}c$ sector, the lower right corner
the $\bar{q}q$ sector. The off-diagonal elements quantify
the mixing.

\begin{figure}
\includegraphics[width=.48\textwidth,clip]{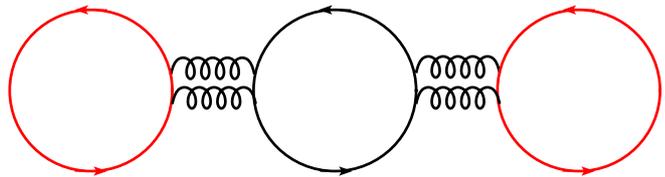}
\caption{The lowest order perturbative QCD graph responsible
for implicit mixing between states
created by $\bar{c}\gamma_5 c$ and by $\bar{q}\gamma_5 q$ operators.
\label{implmix}}
\end{figure}

Note that already in the pure charmonium sector
``implicit'' mixing will occur, due
to intermediate light quark loops in the disconnected part,
see Fig.~\ref{implmix}, or even intermediate glueball
states. If the explicit mixing encoded in the ${\mathcal O}(\lambda)$
off-diagonal elements of
the mixing matrix is small then we will not be able to
resolve the ${\mathcal O}(\lambda^2)$ decay of
$\bar{c}c$ states into states dominated by $\bar{q}q$ within any
sensible Euclidean time distances.
This further justifies and motivates our mixing matrix approach.

Our strategy is as follows. We first determine the eigenvalues and eigenstates
of the three by three submatrices within each of the flavour sectors,
separately, in order to obtain an ``unperturbed'' approximation to the
spectrum. We will then compare the spectrum and the eigenvector components
of this reference point to the situation with the mixing elements switched on.

The all-to-all propagator estimates for both, the charm
and light disconnected loops have been improved by the HPE, obdSSP
and colour partitioning, see Sec.~\ref{Se:a2aprop}.
For the light quark propagators, in addition
the TSM~\cite{Collins:2007mh,Bali:2009hu}
with $n_t=25$ has been applied. We analyse ensemble \mycirc{1},
see Table~\ref{confdetail_tab}, with a pseudoscalar mass of about
1~GeV. At this heavy mass we find an $\eta$-$\pi$ mass splitting
of only $52(13)$~MeV. 
Within the diagonal three by three submatrices we find the
disconnected charmonium loops to be very noisy.
Since the statistical errors are bigger than the expected
splitting of a few MeV we ignore these contributions.
If we were to detect significant off-diagonal
contributions to the full correlation matrix
then of course we would have to revisit this issue at a later stage.

The masses of the light $\eta$ meson and of its first
radial excitation\footnote{This should not be confused with the pseudoscalar
flavour-singlet meson
in the $n_{\mathrm{F}}=2+1$ theory.}
$\eta'$ can be extracted from the largest two eigenvalues
of the submatrix containing the light quark creation and
annihilation operators while
the $\eta_c$ and $\eta_c'$ masses can be approximated from
the $\bar{c}c$ sector. We find a diagonalization of the full six by six
matrix to be numerically unstable and hence restrict
ourselves to the basis
of the states $(\bar{c}\gamma_5{c})_{10}$, $(\bar{c}\gamma_5{c})_{80}$,
$(\bar{q}\gamma_5{q})_{5}$ and $(\bar{q}\gamma_5{q})_{40}$ for our
subsequent full-fledged mixing analysis.

\begin{figure}
\includegraphics[height=.48\textwidth,clip,angle=270]{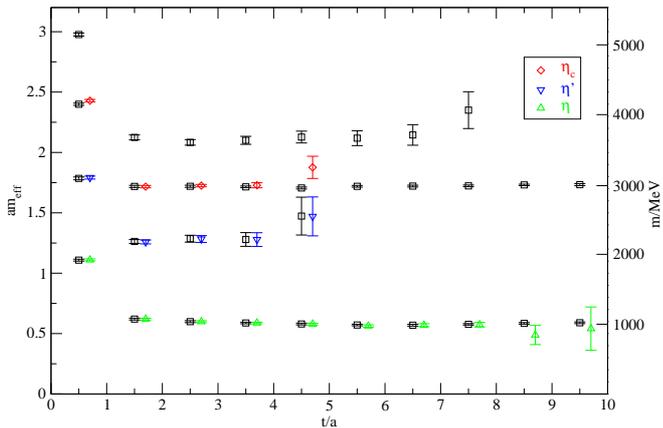}
\caption{Effective masses from the eigenvalues of the full matrix. As a reference point the effective masses from the (unmixed)
submatrices are plotted too (black squares).
\label{eta_eta_c_em_full}}
\end{figure} 

In Fig.~\ref{eta_eta_c_em_full} we display the effective masses of the
$\eta$, $\eta'$, $\eta_c$ and $\eta_c'$ states, with the off-diagonal
matrix elements switched off (squares). These are compared
to the lowest three effective masses obtained from
the four by four matrix with the
mixing switched on, as explained above.
We shift the latter effective masses slightly to the right.
No relevant deviations can be seen.

\begin{figure}
\includegraphics[height=.48\textwidth,clip,angle=270]{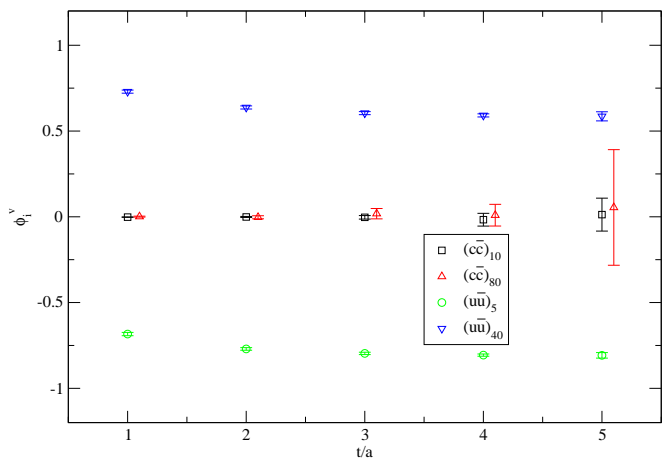}
\caption{Eigenvector components of $\eta$ in the full basis.
\label{eta_eta_c_evec1}}
\end{figure} 

\begin{figure}
\includegraphics[height=.48\textwidth,clip,angle=270]{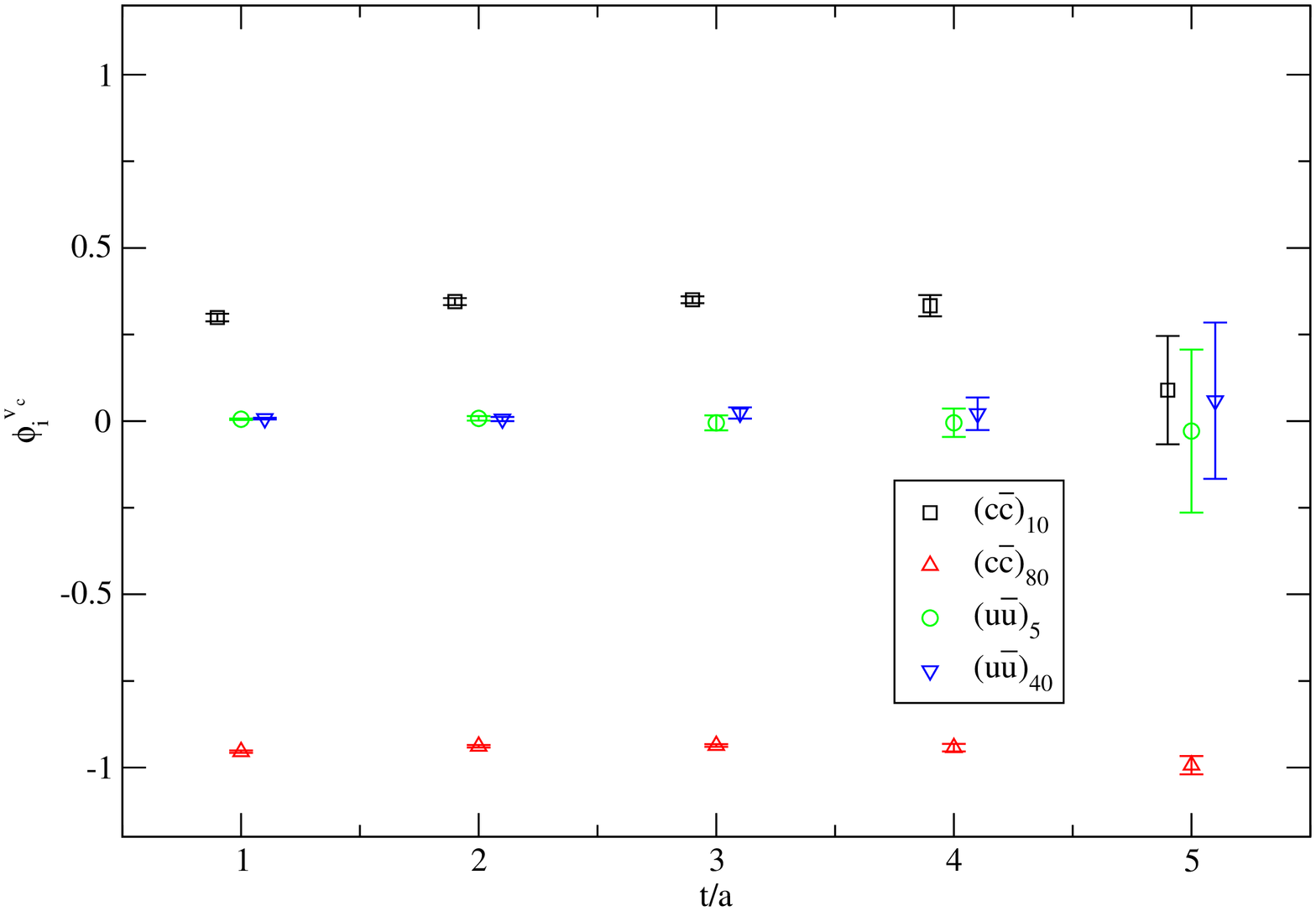}
\caption{Eigenvector components of $\eta_c$ in the full basis.
\label{eta_eta_c_evec2}}
\end{figure} 

\begin{table}
\caption{Fitted eigenvector components of the $\eta$ and $\eta_c$
states from the diagonalization of the full matrix.\label{etacetaptab}}
\begin{center}
\begin{ruledtabular}
\begin{tabular}{ccccc}
  & $(c\bar{c})_{10}$ & $(c\bar{c})_{80}$ & $(q\bar{q})_5$ & $(q\bar{q})_{40}$ \\
 \hline
 $\eta$ & -0.017(37) & 0.009(63) & -0.806(1) & 0.591(9) \\
 $\eta_c$ & 0.333(30) & 0.943(11) & ~0.000(41) & 0.021(47)
\end{tabular}
\end{ruledtabular}
\end{center}
\end{table}

The mixing can be studied in more
detail by investigating the respective effective eigenvector components
that are defined in Eq.~(\ref{eq:overl}) where
$U$ diagonalizes $C(t_0)$ that has the eigenvalues $d_i^2$.
We display these effective overlaps for the ground state $\eta$ meson in
Fig.~\ref{eta_eta_c_evec1} and for the $\eta_c$
in Fig.~\ref{eta_eta_c_evec2}.
The fitted components are displayed in Table~\ref{etacetaptab}.
Indeed, the $\eta$ does not receive any statistically
relevant $\bar{c}c$ contribution and vice versa.
The summed probability to find the $\eta$ meson in a
state that can be created by the $\bar{c}c$ operators
amounts to $(4\pm 25)\cdot 10^{-4}$ and to find
the $\eta_c$ in a state created by $\bar{q}q$
to $(4\pm 22)\cdot 10^{-4}$. These tiny upper limits on
possible mixing effects also render a relevant coupling of the $\eta_c$
state to the pseudoscalar glueball extremely unlikely since this
glueball can appear as an intermediate state
in diagrams of the type depicted in Fig.~\ref{ccuumix_fig}.

Obviously, the energy shift from explicitly admitting
$\bar{c}c$ annihilation and light quark creation cannot
significantly differ from zero either.
We find a mass difference,
$m_{\eta_c}^{\mathrm{mixed}}-m_{\eta_c}^{\mathrm{unmixed}}=11(24)$~MeV.
After this demonstration of the feasibility of such studies
we wish to further reduce the statistical errors and to vary the
light quark mass in the near future.
We address contributions from higher Fock states that may be relevant, e.g.\
for the $\eta_c'$ state in the following section.

\section{Mixing between $\mathbf{\bar{c}c}$ and
$\mathbf{D\overline{D}}$ molecular or tetraquark states}
\label{sec:moleculemix}
Charmonia can decay into pairs of (excited) $D$ and $\overline{D}$
mesons if their masses are above the allowed decay
thresholds.
Charmonia near these thresholds may however also contain significant
Fock admixtures of $D\overline{D}$ molecules, see e.g.\
Ref.~\cite{Eichten:2005ga},
or of $\bar{c}\bar{q}qc$ tetraquarks. 
We will study these effects in three different $J^{PC}$ channels,
$0^{-+}$, $1^{--}$ and $1^{++}$. The first two channels
are interesting with respect to the experimental overpopulation
of the vector channel and the fact that the 
$\Psi(2S)$-$\eta_c'$
finestructure splitting is very small, compared to
that of the ground states, see the
discussion of Secs.~\ref{sec:discuss} and \ref{sec:etamix} above. The $1^{++}$
is phenomenologically relevant to disentangle the nature of
the $X(3872)$ state.

We will address the question of higher Fock state
contributions to the spectrum by creating and destroying states
employing both, the $\bar{c}c$ operators corresponding to
$\eta_c$, $J/\Psi$ and $\chi_{c1}$ charmonia as well as the
four-quark operators corresponding to
$D_1\overline{D}^*$ in $J=0$, $D_1\overline{D}$ and
$D^*\overline{D}$ molecules, respectively.
The analysis method is exactly the same as outlined in
Sec.~\ref{sec:etamix}. However, in the present situation the dependence of
the mixing strength of Eq.~(\ref{eq:mix}) (replacing
$\bar{c}c\mapsto \bar{c}q\bar{q}c$,
$\bar{q}q\mapsto \bar{c}c$)
on the light quark mass is evident: again, with
decreasing light quark masses the numerator is likely
to increase, however, the denominator will decrease as the
energy gaps between open charm states and the first radial charmonium
excitations become smaller. Therefore, we analyse ensemble \mycirc{3}
(see Table~\ref{confdetail_tab})
that, with a light pseudoscalar mass of about 280~MeV, is closest to
the physical point. Note that with the product
$m_{\mathrm{PS}}L\approx 2.6$ this $L\approx 1.84$~fm lattice
is quite small so that in particular for radial excitations we
may expect significant finite size effects.

\begin{table}
\caption{$\Gamma$ structures of meson and molecule
interpolating fields, see Eqs.~(\protect\ref{eq:meson})
and (\protect\ref{eq:molecule}).\label{molecule_tab}}
\begin{center}
\begin{ruledtabular}
\begin{tabular}{ccccc}
 $J^{PC}$ & $\Gamma_{\mathrm{M}}$ & $\Gamma_{\mathrm{Y}}^1$ &
$\Gamma_{\mathrm{Y}}^2$ & $s$ \\
 \hline
 $0^{-+}$ & $\gamma_5$ & $\gamma_i$ & $\gamma_i\gamma_5$ & $0$ \\
 $1^{--}$ & $\gamma_i$ & $\gamma_5$ & $\gamma_i\gamma_5$ & $1$ \\
 $1^{++}$ & $\gamma_i\gamma_5$ & $\gamma_5$ & $\gamma_i$ & $1$
\end{tabular}
\end{ruledtabular}
\end{center}
\end{table}

We start from a six dimensional operator basis containing
three $\bar{c}c$ and three molecular interpolators,
differing by their Wuppertal smearing levels.
We label these as p(oint), n(arrow) and w(ide).
The generic form of the meson operators
centred at the spacetime position
$x$ reads,
\begin{equation}
\label{eq:meson}
M_x=(\bar{c}\Gamma_{\mathrm{M}}c)_x \, ,
\end{equation}
where for readability we omit the smearing functions.
Within the molecular operators we allow for a spatial separation
$\mathbf{r}$,
\begin{align}
Y_x(\mathbf{r})=&
\frac{1}{\sqrt{2}}\left[(\bar{q}\Gamma_{\mathrm{Y}}^1c)_x(\bar{c}
\Gamma_{\mathrm{Y}}^2q)_{x+\mathbf{r}}\right.\nonumber\\\label{eq:molecule}
 &+ (-)^s \left.(\bar{c}\Gamma_{\mathrm{Y}}^1q)_x(\bar{q}
\Gamma_{\mathrm{Y}}^2c)_{x+\mathbf{r}}\right] \, .
\end{align}
The $\Gamma$ structures and $s\in\mathbb{N}_0$ values for the
states of interest are displayed in
Table~\ref{molecule_tab}, see also Ref.~\cite{Chiu:2006ge}.
Note that in this exploratory study we restrict ourselves
to operators with molecular contractions and ignore the possibility
of arranging the quarks into tetraquark-like
$\bar{c}\bar{q}$ and $qc$ diquark-antidiquark structures.

\begin{figure}
\includegraphics[width=.48\textwidth,clip]{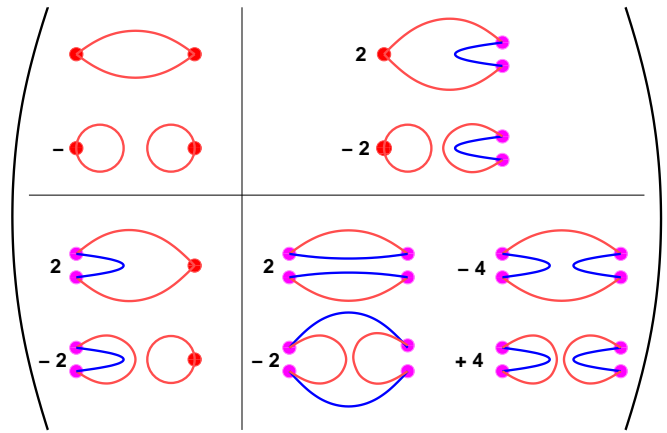}
\caption{Correlation matrix for the mixing of
charmonia with $D\overline{D}$ molecules. Red lines represent
charm quark propagators, blue lines light quarks.
\label{mixingmatrix_fig}}
\end{figure} 

In Fig.~\ref{mixingmatrix_fig} we sketch the structure
of the mixing matrix. The different smearing levels are again
omitted for clarity. Red lines represent charm quark propagators
and blue lines light quark propagators.
The prefactors are due to the $n_{\mathrm{F}}=2$
mass degenerate light sea quark flavours. The upper left
corner contains the $\bar{c}c$, the lower right corner the
molecular sector. The off-diagonal elements encode
explicit mixing. In our calculation
we omit the charm annihilation diagrams of the second lines
within each of the correlation matrix sectors;
based on
our experience of Sec.~\ref{sec:etamix} above we deem
these numerically irrelevant.
A similar matrix was constructed, e.g.\ in
Refs.~\cite{Aoki:2007rd,Prelovsek:2010kg} in order to
investigate the $\rho$ meson decay width and light tetraquark
states, respectively. Note that all the depicted diagrams
that include light quark propagators include more than one
quark line contraction since for $\mathbf{r}\neq\mathbf{0}$
Eq.~(\ref{eq:molecule}) contains two terms.

The spatial separation within the molecular operators was
tuned to maximize the correlation function of
the molecular sector. This
led us to employ the on-axis separation
$r=4a\approx 0.3\,$fm. After some experiments we decided
to employ point-to-all propagators for all diagrams, with the exception
of the top right diagram within the molecule-to-molecule sector,
see Fig.~\ref{mixingmatrix_fig}. This necessitates to implement
a light all-to-all propagator at the sink.
For this purpose
we generated the equivalent of
100 complex $\mathbb{Z}_2$ stochastic estimates, employing
the obdSSP, HPE and TSM methods.

\begin{figure}
\includegraphics[height=.48\textwidth,clip,angle=270]{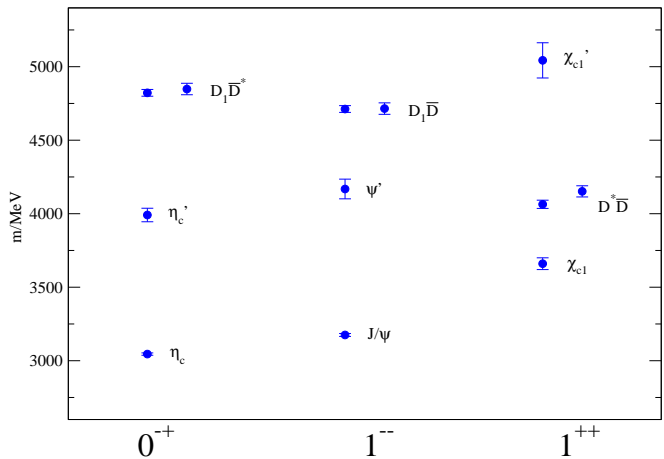}
\caption{Mass spectrum from separately diagonalizing the
submatrices within the different sectors. The right points
for the $D\overline{D}$ states correspond to the sums of non-interacting
$D$ mesons, the left points to interacting $I=0$ $D$ and $\overline{D}$ mesons.
\label{spectrum_fig}}
\end{figure} 

\begin{table}
\caption{Mass spectrum in MeV, neglecting the mixing between the two- and
four-quark sectors.\label{tab:masses}}
\begin{center}
\begin{ruledtabular}
\begin{tabular}{cccc}
$J^{PC}=$&$0^{-+}$&$1^{--}$&$1^{++}$\\\hline
$\bar{c}c$ ground state                      &3045~(8)&3175(10)&3660~(40)\\
$\bar{c}c$ first excitation                  &3991(46)&4168(67)&5043(120)\\
$D_{(1)}+\overline{D}^{(*)}$                 &4848(39)&4715(39)&4152~(38)\\
$D_{(1)}\overline{D}^{(*)}$ molecule         &4822(23)&4712(23)&4064~(28)
\end{tabular}
\end{ruledtabular}
\end{center}
\end{table}

We first  diagonalize the submatrices separately within the
$\bar{c}c$ and molecular sectors to obtain a reference
spectrum. This provides us with up to four reliable eigenvalues,
two within each sector. However, the excited molecular channels
are quite noisy so that in these cases we are only able to extract acceptable
plateaus for the ground states.
The remaining three states within each of the $J^{PC}$ channels
are plotted in Fig.~\ref{spectrum_fig}. Next to the isospin\footnote{Note that
the $I=1$ channel has been studied by Liu~\cite{Liu:2009zz}.}
$I=0$ molecular masses we also display the sums of the masses of
the corresponding individual $D$ and anti-$D$ mesons.
The resulting masses are also displayed in Table~\ref{tab:masses}
where the errors are statistical only. Due to the finite volume
the radially excited $S$-wave states are significantly higher
than the corresponding experimental masses and the excited $P$-wave
suffers even more from this effect, being by almost 1~GeV heavier than
the corresponding $D^*\overline{D}$ molecule that is already
heavier than in the real world due to the unphysically large
light quark mass.

Within errors of about 30~MeV we do not find any significant
mass differences between molecules and their
open charm constituent mesons in the pseudoscalar and vector channels.
Of particular interest is the mass of the $1^{++}$ molecule that is by
almost 200 MeV heavier than the $X(3872)$. However,
this can easily be attributed to the light quark mass since
the light pseudoscalar is still by 140~MeV heavier than the
physical one.  
We find a significant binding of this
axialvector molecule, $m_{D^*\overline{D}}-(m_{D^*}+m_D)=88(26)$~MeV. 
There will be some volume and light quark mass dependence of this
value that needs to be studied. Note that this binding energy
is much bigger
than the mass differences between the experimental
$X(3872)$
of a fraction of an MeV with respect to electrically neutral open
charm states and of a few MeV with respect to charged $D$ and $D^*$
mesons. However, on a qualitative level the increased attraction
deserves some attention.

\begin{figure}
\includegraphics[height=.48\textwidth,clip,angle=270]{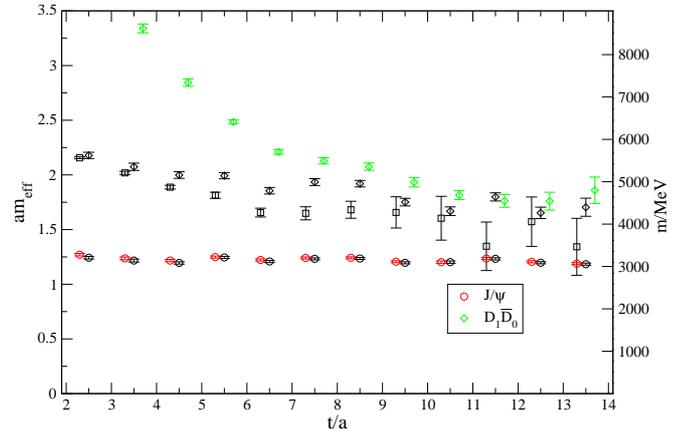}
\caption{Effective masses from the eigenvalues of the full matrix in the $1^{--}$ channel. As a reference point the effective masses from the submatrices are plotted too (black symbols).
\label{vecevfull}}
\end{figure} 

\begin{figure}
\includegraphics[width=.48\textwidth,clip]{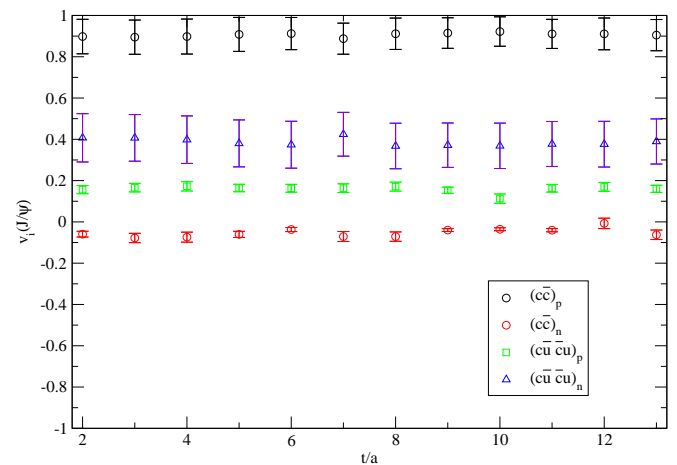}
\caption{Eigenvector components of $J/\Psi$ in the full basis.
\label{vecevec1}}
\end{figure} 

\begin{figure}
\includegraphics[width=.48\textwidth,clip]{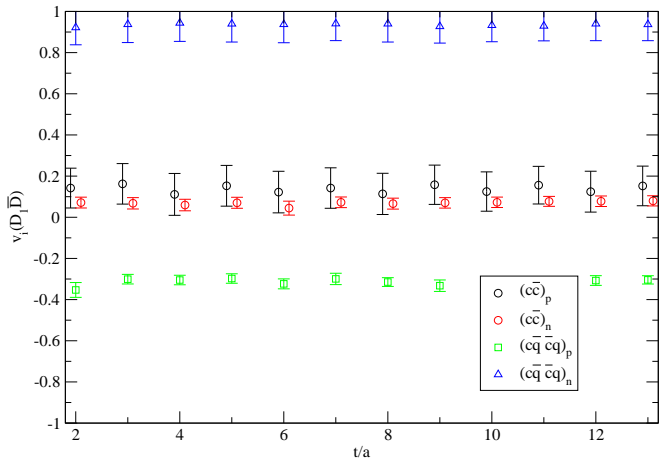}
\caption{Eigenvector components of $D\overline{D}_1$ in the full basis.
\label{vecevec2}}
\end{figure} 

We base our fully fledged mixing study on a four by four submatrix with
the operator basis
$M_{\mathrm{p}}, M_{\mathrm{n}}, Y_{\mathrm{p}}$
and $Y_{\mathrm{n}}$. The normalization time
is $t_0=2a$ for all channels.
We start the discussion with the vector state and
display the resulting lowest two effective masses, together
with the unmixed reference masses in Fig.~\ref{vecevfull}.
The corresponding effective overlaps Eq.~(\ref{eq:overl}) 
are plotted in Fig.~\ref{vecevec1} for the $J/\Psi$
and in Fig.~\ref{vecevec2} for the $D_1\overline{D}$ molecule.
The $J/\Psi$ receives the dominant contributions from the $\bar{c}c$
sector. However, the molecular configurations contribute
significantly too. In contrast, the $D_1\overline{D}$ state only contains
a small (but non-vanishing) $\bar{c}c$ admixture. This is very similar
to the observation of Ref.~\cite{Bali:2005fu} that the ground state
potential between two static sources $Q$ and $\overline{Q}$
receives a significant light quark contribution also
for distances much smaller than the string breaking distance
while its $\overline{Q}q\bar{q}Q$ excitation contains almost no
$\overline{Q}Q$ component.
We obtain the same qualitative picture for the pseudoscalar.

\begin{figure}
\includegraphics[height=.48\textwidth,clip,angle=270]{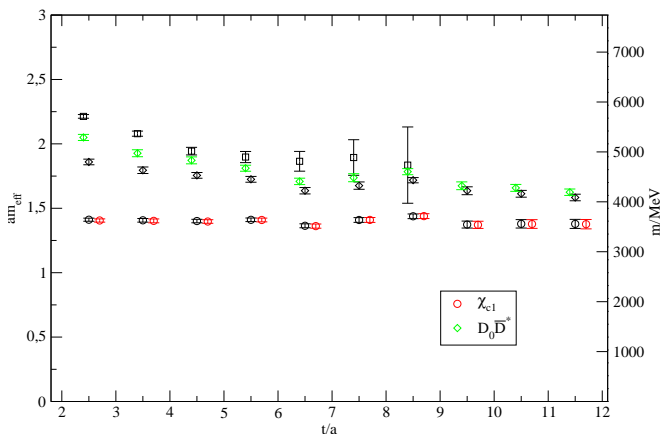}
\caption{Effective masses of the eigenvalues of the full matrix in the $1^{++}$ channel. As a reference point the effective masses from the submatrices are plotted too (black symbols).
\label{pwaveevfull}}
\end{figure} 

\begin{figure}
\includegraphics[width=.48\textwidth,clip]{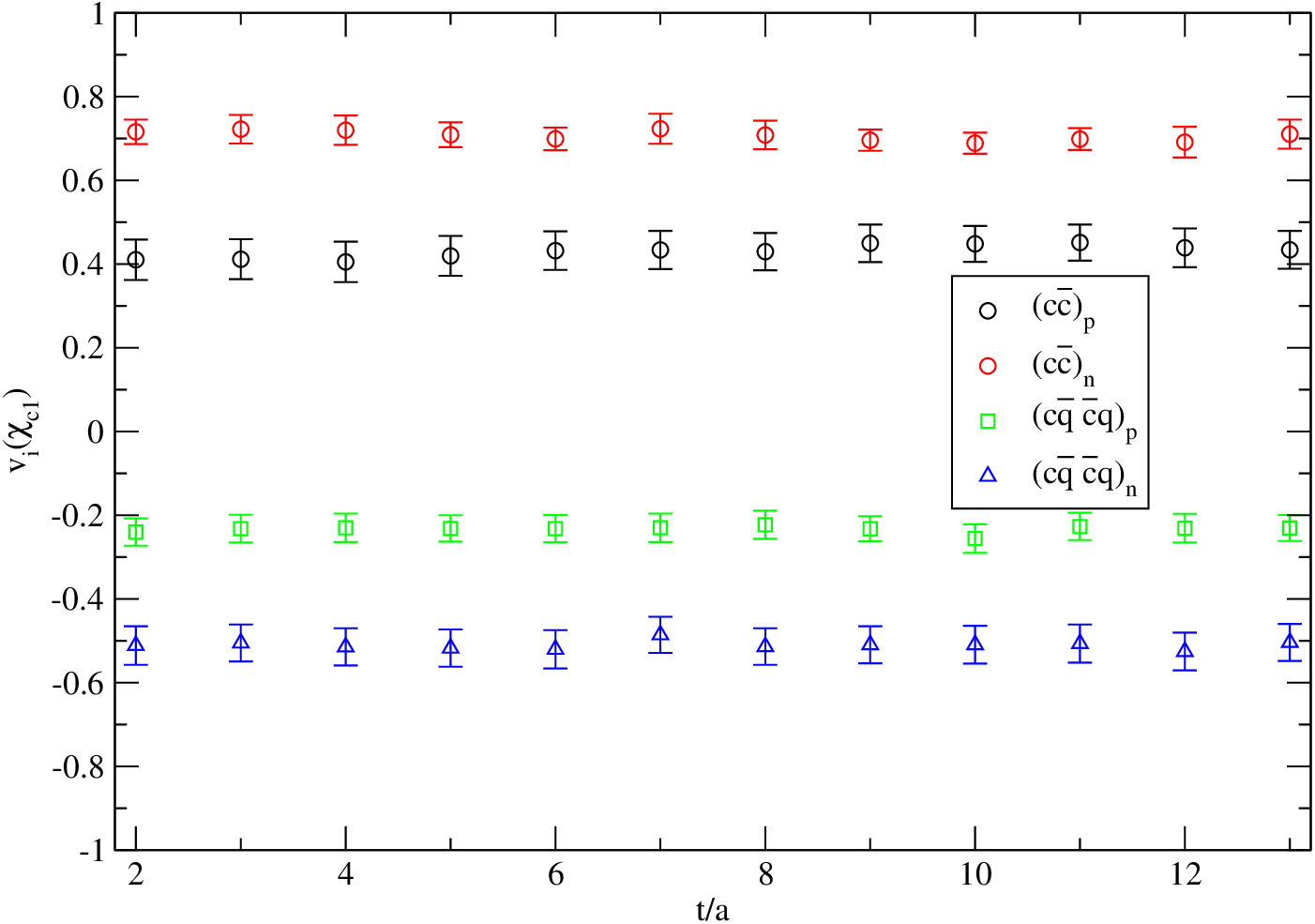}
\caption{Eigenvector components of $\chi_{c1}$ in the full basis.
\label{pwaveevec1}}
\end{figure} 

\begin{figure}
\includegraphics[width=.48\textwidth,clip]{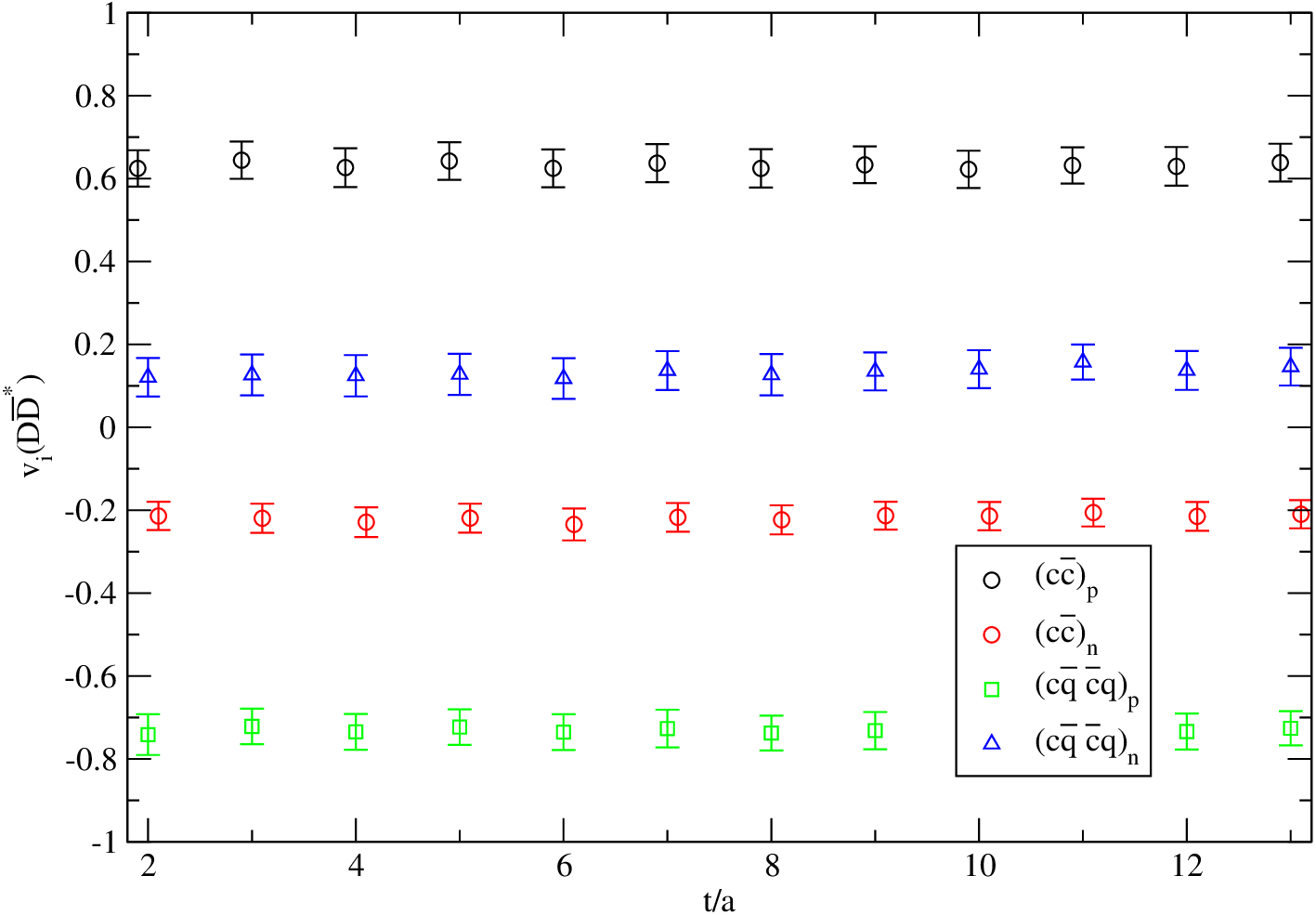}
\caption{Eigenvector components of $D\overline{D}^*$ in the full basis.
\label{pwaveevec2}}
\end{figure} 

\begin{figure}
\includegraphics[width=.48\textwidth,clip]{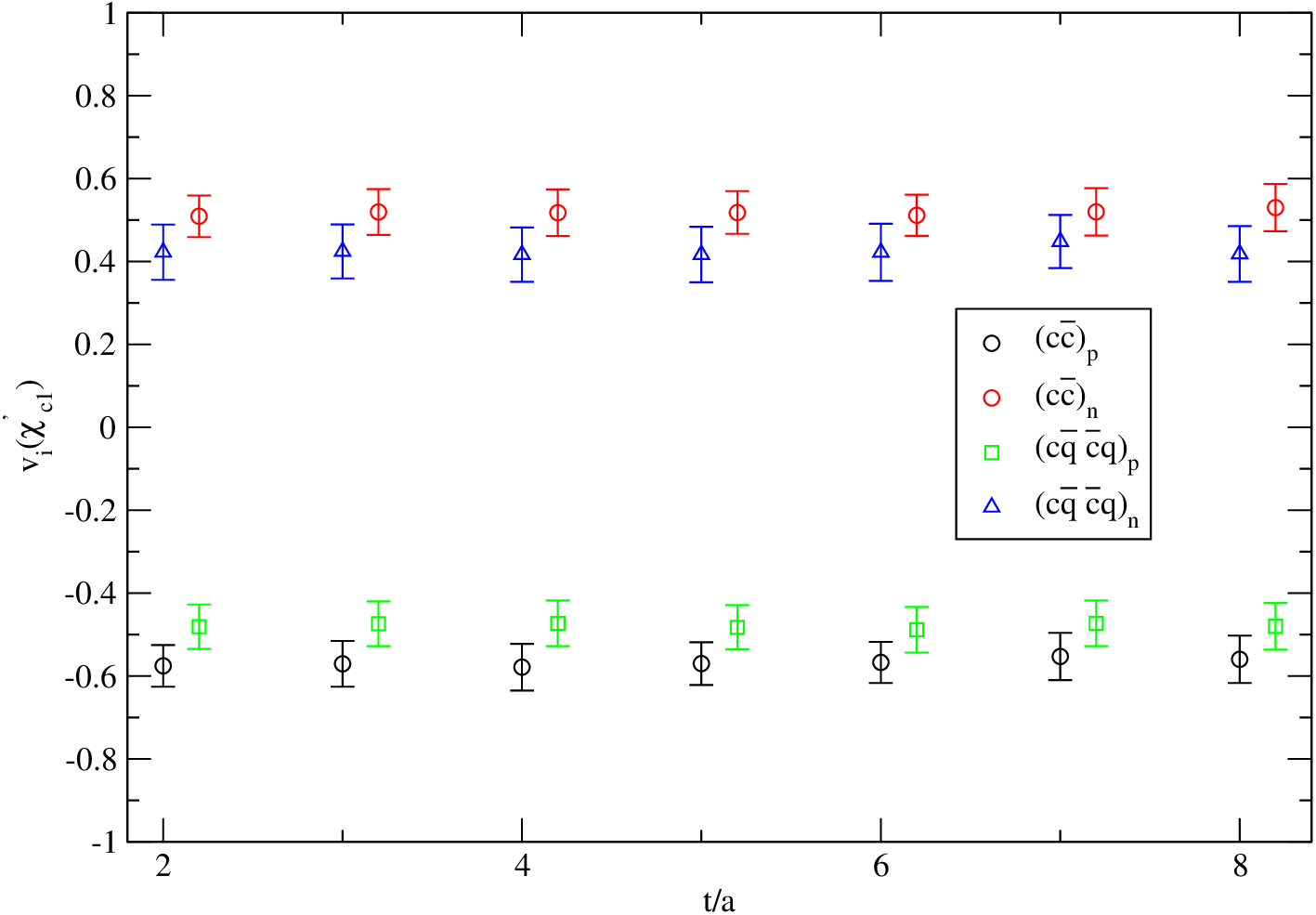}
\caption{Eigenvector components of $\chi_{c1}'$ in the full basis.
\label{pwaveevec3}}
\end{figure} 

\begin{table}
\caption{Eigenvector components in the full basis.\label{eigenveccomp_tab}}
\begin{center}
\begin{ruledtabular}
\begin{tabular}{ccccc}
 & $(\bar{c}c)_{\mathrm{p}}$ & $(\bar{c}c)_{\mathrm{n}}$ &
$(\bar{c}q\bar{q}c)_{\mathrm{p}}$ &
$(\bar{c}q\bar{q}c)_{\mathrm{n}}$ \\\hline
 $\eta_c$ & ~0.87~(5) & -0.03~(2) & -0.02~(1) & -0.50~(7) \\
 $D_1\overline{D}^*$ & ~0.14~(2) & ~0.02~(2) & -0.95(15) & ~0.29~(4) \\
 \hline
 $J/\Psi$ & ~0.91~(7) & -0.05~(2) & ~0.16~(2) & ~0.38(11) \\
 $D_1\overline{D}$ & ~0.14(11) & ~0.07~(2) & -0.32~(2)  & ~0.93~(8) \\
 \hline
 $\chi_{c1}$ & ~0.41~(4) & ~0.72~(3) & -0.23~(3)  & -0.51~(4) \\
 $D\overline{D}^*$ & ~0.63~(4) & -0.23~(3) & -0.73~(4) & ~0.12~(3) \\
 $\chi_{c1}'$ & -0.55~(6) & ~0.53~(5) & -0.49~(5)  & ~0.41~(6)
\end{tabular}
\end{ruledtabular}
\end{center}
\end{table}

For the axialvector the situation is different. We display the mixed and
unmixed effective masses in Fig.~\ref{pwaveevfull}. Note that in this case
the mixed $D^*\overline{D}$
mass slightly increases, relative to the unmixed result.
The corresponding effective overlaps are displayed in Figs.~\ref{pwaveevec1},
\ref{pwaveevec2} and \ref{pwaveevec3}. The fitted results on the
operator overlaps are summarized in
Table~\ref{eigenveccomp_tab}. The relative probability of creating
the $\chi_{c1}$ by a four-quark operator is 0.29(5). For the
first excitation that we identify as a $D^*\overline{D}$ molecule
it is  0.53(7) and for the $\chi'_{c1}$ it is 0.36(10): all these states
appear to undergo strong mixing.

This casts doubt onto the validity of the mixing
model Eq.~(\ref{eq:mix}). Setting this problem aside for the moment,
we define unmixed $\chi_{c1}$ wavefunctions, projecting onto
the $\bar{c}c$ components of the space spanned by our operator basis,
\begin{equation}
|\chi_{c1}\rangle^{\mathrm{un}}=
d^{\chi_{c1}}_{(\bar{c}c)_\mathrm{p}}|(\bar{c}c)_{\mathrm{p}}\rangle
+d^{\chi_{c1}}_{(\bar{c}c)_\mathrm{n}}|(\bar{c}c)_{\mathrm{n}}\rangle,
\end{equation}
and similarly for the excitation, $\chi_{c1}'$, while for
the $D^*\overline{D}$ state we can define the
projection onto its four-quark components, $|D^*\overline{D}\rangle^{\mathrm{un}}$.
Of physical interest are the overlaps between these idealized
unmixed states and the respective physical states.
These can be obtained by computing,
\begin{equation}
\langle D^*\overline{D}|\chi_{c1}\rangle^{\mathrm{un}}=
d^{\chi_{c1}}_{(\bar{c}c)_\mathrm{p}}
d^{D^*\overline{D}}_{(\bar{c}c)_\mathrm{p}}+
d^{\chi_{c1}}_{(\bar{c}c)_\mathrm{n}}
d^{D^*\overline{D}}_{(\bar{c}c)_\mathrm{n}}\,.
\end{equation}
The resulting probabilities read as follows,
\begin{align}
|\langle D^*\overline{D}|\chi_{c1}\rangle^{\mathrm{un}}|^2&=
0.01(1)\,,\quad
|\langle\chi_{c1}| D^*\overline{D}\rangle^{\mathrm{un}}|^2=
0.01(1)\,,\nonumber\\
|\langle \chi_{c1}'|\chi_{c1}\rangle^{\mathrm{un}}|^2&=
|\langle \chi_{c1}|\chi_{c1}'\rangle^{\mathrm{un}}|^2=
0.01(1)\,,\\
|\langle D^*\overline{D}|\chi_{c1}'\rangle^{\mathrm{un}}|^2&=
0.22(5)\,,\quad
|\langle\chi_{c1}| D^*\overline{D}\rangle^{\mathrm{un}}|^2
=0.41(7)\nonumber
\end{align}
while for the normalizations we obtain,
\begin{align}
|\langle\chi_{c1}|\chi_{c1}\rangle^{\mathrm{un}}|^2&=0.47(7)\,,\quad
|\langle D^*\overline{D}| D^*\overline{D}\rangle^{\mathrm{un}}|^2
=0.30(6)\,,\nonumber\\
|\langle\chi_{c1}'|\chi_{c1}'\rangle^{\mathrm{un}}|^2&=0.34(5)\,.
\end{align}
Due to cancellations the groundstate axialvector charmonium
does not actively participate in the mixing while
the radial excitation and the molecular state strongly
mix with each other. However, the truncation of the mixing model
at $\mathcal{O}(\lambda)$
is only justifiable on a qualitative level,
as is obvious from $|\langle\chi_{c1}|\chi_{c1}\rangle^{\mathrm{un}}|^2< 1$.

\section{Summary and outlook}
\label{sec:sum}
We introduced the tools necessary to study the mixing of
standard charmonium states with states created by four-quark
operators and with light quark flavour-singlet states.
Of particular importance was the use of the variational generalized
eigenvalue method as well as 
of improved stochastic all-to-all
propagator methods. We introduced the staggered spin partitioning (SSP)
and the recursive noise subtraction (RNS) methods
(see also Ref.~\cite{Ehmann:2009ki}). We also
made use of the hopping parameter expansion (HPE)
subtraction method~\cite{Thron:1997iy} and of
the truncated solver method (TSM)~\cite{Collins:2007mh,Bali:2009hu}.

Our spin-averaged charmonium spectrum agrees fairly well with
the experimental data. However, due to our unphysically
heavy sea quark masses with light pseudoscalar masses
ranging from 1~GeV
down to 280~MeV, due to the fact that we are simulating
with $n_{\mathrm{F}}=2$ sea quarks only and possibly
due to the use of somewhat coarse
lattice spacings of $a\approx 0.115$~fm and $a\approx 0.077$~fm,
we significantly underestimate finestructure splittings.
The ratio of the $2S$ finestructure splitting over the
$1S$ splitting from which we expect some of the systematics
to cancel comes out significantly larger than in experiment.
One reason may be a distortion of the radial excitations
due to their proximity to open charm thresholds which lie
higher in our simulations than in the real world.

The lowest spin-exotic state is a $1^{-+}$ vector with a mass
of 4.15(5)~GeV where the error is statistical only. The next highest
such state can be found at 4.61(22)~GeV with quantum numbers $2^{+-}$.
We interpret the small mass differences that we find
with respect to radial excitations
in these sectors as evidence of a charm-anticharm-gluon hybrid nature
of these states. In other $J^{PC}$ sectors we do not see such evidence.
At least for masses below 3.8~GeV we do not detect any mixing between
$S$- and $D$-waves, indicating that $L$ is a relatively good quantum
number for this mass range. This conclusion is also supported by examining
the spatial structure of the optimized creation operators that
we employ.

We realize that, to exclude mass shifts that are due to the flavour-singlet
nature of charmonium states, it is not sufficient just to
incorporate quark line disconnected charm annihilation diagrams but
that we also have to consider mixing with light flavour-singlet
states. However, within errors of less than one per mille,
we do not detect any light quark contributions to charmonium wavefunctions
and vice versa.
Moreover, within statistical errors of 24 MeV we do not find
any significant mass shift:
$m_{\eta_c}^{\mathrm{mixed}}-m_{\eta_c}^{\mathrm{unmixed}}=11(24)$~MeV.
Clearly, in future studies we will aim at reducing this error.

We then moved on to investigate the binding between
pairs of $D$ and anti-$D$ mesons in the pseudoscalar,
vector and axialvector sectors. Only the axialvector channel was
clearly attractive, however, we emphasize that the
volume scaling still needs to be investigated for definite
conclusions. Subsequently, the mixing between
isoscalar charmonium
states created by two- and four-quark operators was
investigated. Within
the vector and pseudoscalar sector, at a light 
quark mass value that is four times as large as the
physical one, these effects exist but they are small.
However, in the axialvector channel the state that is
dominated by the radial
charmonium excitation strongly couples to the
$D^*\overline{D}$ molecular state and vice versa.
This is very interesting in view of the nature of the experimental
$X(3872)$ state.

We plan to apply the methods that we developed and tested here
in high precision studies of charmonium states with
$n_{\mathrm{F}} =2+1$ sea quarks of different masses
on various volumes and lattice spacings, within a large
collaboration. First results
of these systematic investigations were presented
at the Lattice 2011 conference~\cite{Bali:2011dc}.
\begin{acknowledgments}
We thank the QCDSF Collaboration for making their configurations
available to us.
This work was supported by the European Union
under Grant Agreement numbers 238353 (ITN STRONGnet)
and 227432 (IA Hadronphysics2) and by the Deutsche
Forschungsgemeinschaft SFB/Transregio 55.
Sara Collins
acknowledges support from the Claussen-Simon-Foundation (Stifterverband
f\"ur die Deutsche Wissenschaft).
The computations were performed on the
(now de-commissioned) QCDOC machine
and on the Athene HPC cluster at the Universit\"at Regensburg,
on the IBM BlueGene/L at the Edinburgh Parallel Computing
Centre and on the IBM 
BlueGene/P (JuGene) of the
J\"ulich Supercomputer Center. We thank the
support staffs of these institutions.
We made extensive use of the Chroma software suite~\cite{Edwards:2004sx}.\\

\end{acknowledgments}

\end{document}